\pdfoutput=1

\documentclass[11pt,twoside,a4paper,cmspaper,final,collab]{cms-tdr}
\def\svnVersion{329654}\def\cmsCernNo{2013-037}\def\cmsCernDate{\today}\def\cmsMessage{Published in the Journal of High Energy Physics as \href{http://dx.doi.org/10.1007/JHEP02(2016)122}{\doi{10.1007/JHEP02(2016)122.}}}

\begin{document}\cmsNoteHeader{B2G-12-009}

\hyphenation{had-ron-i-za-tion}
\hyphenation{cal-or-i-me-ter}
\hyphenation{de-vices}
\RCS$Revision: 323958 $
\RCS$HeadURL: svn+ssh://svn.cern.ch/reps/tdr2/papers/B2G-12-009/trunk/B2G-12-009.tex $
\RCS$Id: B2G-12-009.tex 323958 2016-02-07 02:55:25Z alverson $
\newlength\cmsFigWidth
\ifthenelse{\boolean{cms@external}}{\setlength\cmsFigWidth{0.85\columnwidth}}{\setlength\cmsFigWidth{0.4\textwidth}}
\ifthenelse{\boolean{cms@external}}{\providecommand{\cmsLeft}{top}}{\providecommand{\cmsLeft}{left}}
\ifthenelse{\boolean{cms@external}}{\providecommand{\cmsRight}{bottom}}{\providecommand{\cmsRight}{right}}
\newcommand{\wpr}{\ensuremath{\PWpr}\xspace}
\newcommand{\wprr}{\ensuremath{\PWpr_{\mathrm{R}}}\xspace}
\newcommand{\wprl}{\ensuremath{\PWpr_{\mathrm{L}}}\xspace}
\newcommand{\wprlr}{\ensuremath{\PWpr_{\mathrm{LR}}}\xspace}
\newcommand{\mtb}{\ensuremath{M_{\PQt \PQb}}\xspace}
\cmsNoteHeader{B2G-12-009}
\title{Search for \texorpdfstring{$ \wpr \to \PQt \PQb $}{W' to tb} in proton-proton collisions at \texorpdfstring{$\sqrt{s} = 8\TeV$}{sqrt(s) = 8 TeV}}

\date{\today}

\abstract{
A search is performed for the production of a massive $\wpr$ boson decaying to a top and a bottom quark. The data analysed correspond to an integrated luminosity of 19.7\fbinv collected with the CMS detector at the LHC in proton-proton collisions at $\sqrt{s}=8\TeV$.  The hadronic decay products of the top quark with high Lorentz boost from the $\wpr$ boson decay are detected as a single top flavoured jet. The use of jet substructure algorithms allows the top quark jet to be distinguished from standard model QCD background. Limits on the production cross section of a right-handed $\wpr$ boson are obtained, together with constraints on the left-handed and right-handed couplings of the $\wpr$ boson to quarks. The production of a right-handed $\wpr$ boson with a mass below 2.02\TeV decaying to a hadronic final state is excluded at 95\% confidence level. This mass limit increases to 2.15\TeV when both hadronic and leptonic decays are considered, and is the most stringent lower mass limit to date in the tb decay mode.
}

\hypersetup{%
pdfauthor={CMS Collaboration},%
pdftitle={Search for W' to tb in proton-proton collisions at sqrt(s) = 8 TeV},%
pdfsubject={CMS},%
pdfkeywords={CMS, physics, W'}}

\maketitle
\section{Introduction}
\label{sec:introduction}

Many extensions of the standard model (SM) predict new massive charged gauge bosons
\cite{doi:10.1146/annurev.nucl.55.090704.151502,PhysRevD.64.035002,PhysRevD.11.566}.
The $\wpr$ boson, for example, is a heavy partner of the SM W gauge boson that could manifest itself in
proton-proton collisions at the CERN LHC.  Searches for a high-mass $\wpr$ boson resonance have been performed at the
Tevatron \cite{PhysRevLett.100.211803,Abazov2011145} and the
LHC \cite{Chatrchyan:2014koa,Khachatryan:2014tva,Khachatryan:2014xja,ATLAS:2012pu,Aad:2013wxa,Aad:2014xea,Aad:2014xra,Aad:2012dm,Khachatryan:2014hpa,Chatrchyan:2013lga}
in the lepton, diboson, and diquark final states.
We present a search based on the   $\PW^{\prime +} \to \PQt \PAQb$ (and charge conjugate) decay.
This decay channel is of particular interest because the SM backgrounds can be greatly reduced compared to those for $\wpr$ decays to light quarks,
and some models predict a stronger $\wpr$ coupling to third generation quarks~\cite{Muller1996dj}.
The hadronic decay channel ($\wpr \to \PQt \PQb \to \PQq \PQq \PQb \PQb$) is presented in detail, along with the
combination with the already published leptonic channel ($\wpr \to \PQt \PQb  \to \ell\nu \PQb \PQb$)~\cite{Chatrchyan:2014koa}.
The combination of these two channels leads to a significant increase in sensitivity.

The most general, lowest dimension effective Lagrangian
that describes the interaction of the $\wpr$ boson with quarks \cite{kfactor} can be written as:
\begin{eqnarray}
{\cal L} = \frac{V_{\mathrm{q}_i\mathrm{q}_j}}{2\sqrt{2}} g_{\mathrm{w}} \mathrm{\overline{q}}_i \gamma_{\mu}
\bigl( a^{\mathrm{R}}_{\mathrm{q}_i\mathrm{q}_j} (1+{\gamma}^5) +a^{\mathrm{L}}_{\mathrm{q}_i\mathrm{q}_j}
(1-{\gamma}^5)\bigr) \wpr \mathrm{q}_j + \mathrm{h.c.},
\label{eqn:Lag}
\end{eqnarray}
where the parameters $a_{\mathrm{q}_i\mathrm{q}_j}^{\mathrm{L}}$ and $a_{\mathrm{q}_i\mathrm{q}_j}^{\mathrm{R}}$ represent the left-handed and right-handed
couplings of the $\wpr$ boson to quarks, $g_{\PW}$ is the SM weak coupling constant, and $V_{\mathrm{q}_i\mathrm{q}_j}$ is the Cabibbo--Kobayashi--Maskawa matrix.  For a SM-like $\wpr$ boson, $a_{\mathrm{q}_i\mathrm{q}_j}^{\mathrm{L}}~=~1$, $a_{\mathrm{q}_i\mathrm{q}_j}^{\mathrm{R}}~=~0$.

Strict cross section upper limits have previously been placed in the case of low mass $\wpr$ signals~\cite{Chatrchyan:2014koa,Abazov2011145}.  At higher masses, specifically $M_{\wpr}
\gtrsim 1.3\TeV$, the top quark is highly energetic.  Because of the Lorentz
boost, the angular separation between the top quark decay products (W boson and b quark) is
small.
The final state particles resulting from the hadronization
of the b quark and the decay of the W boson into light quarks usually overlap,
resulting in a single jet with top flavour, the ``top quark jet", or t jet.
Dedicated methods, applied to resolve the substructure of this
t jet, enable background processes to be strongly suppressed.
We apply b jet identification algorithms (b tagging) to the b jet from the $\wpr$ decay in order to further reduce the SM background.

We reconstruct the $\wpr$ boson mass as the invariant mass of the top and bottom quarks ($\mtb$),
and use the $\mtb$ distribution to derive limits on the
production cross section of the $\wpr$ boson. We also obtain limits
on the couplings of the $\wpr$ boson to quarks, and present the $\wpr$ production
cross section limits obtained from a combination of hadronic and
leptonic decay channels.
The leptonic decay channel alone excludes a $\wpr$ boson with a mass less than $2.05\TeV$~\cite{Chatrchyan:2014koa},
which prior to this paper was the most restrictive limit obtained to date in the tb final state.  The combination of the
leptonic and hadronic channels makes it possible to extend this limit.

\section{The CMS detector}
\label{sec:cms_detector}
The central feature of the CMS apparatus~\cite{Chatrchyan:2008zzk} is a superconducting solenoid of 6\unit{m} internal diameter,
providing a magnetic field of 3.8\unit{T}. Within the field volume are a silicon pixel and strip tracker,
a lead tungstate crystal electromagnetic calorimeter (ECAL), and a brass and scintillator hadron calorimeter (HCAL), each composed of a barrel
and two endcap sections.   Forward calorimeters extend the pseudorapidity coverage provided by the barrel and endcap detectors.
Muons are measured in gas-ionization detectors embedded in the steel flux-return yoke outside the solenoid.

The silicon tracker detects charged particles within the pseudorapidity range $\abs{\eta}< 2.5$.
It consists of 1440 silicon pixel and 15\,148 silicon strip detector modules and is located in the field of the superconducting solenoid.
For non-isolated particles with $1 < \pt < 10\GeV$ and $\abs{\eta} < 1.4$, the track resolutions are typically 1.5\% in $\pt$ and 25--90 (45--150)~$\mum$ in
the transverse (longitudinal) impact parameter \cite{TRK-11-001}.  Non-isolated particle tracks are of particular importance to this analysis, as they
are typically the constituents of jets, and are found within the detector barrel acceptance with an efficiency larger than 90\%.

In the region $\abs{ \eta }< 1.74$, the HCAL towers have widths of 0.087 in pseudorapidity and 0.087 radians in azimuth ($\phi$). In the $\eta$-$\phi$ plane, and for $\abs{\eta}< 1.48$, the
HCAL towers map on to $5 \times 5$ ECAL crystal arrays to form calorimeter towers projecting radially outwards from close to the nominal interaction point. At larger values of $\abs{ \eta }$, the size of the ECAL and HCAL towers increases and the matching ECAL arrays contain fewer crystals. Within each tower, the energy deposits in ECAL and HCAL towers are summed to define the calorimeter tower energies, which are subsequently used to provide the energies and directions of hadronic jets.

 A more detailed description of the CMS detector, together with a definition of the coordinate system used and the relevant
kinematic variables, can be found in Ref.~\cite{Chatrchyan:2008zzk}.

\section{Event reconstruction}
\label{sec:EVREC}

The particle-flow (PF) event algorithm~\cite{CMS-PAS-PFT-09-001,CMS-PAS-PFT-10-001}  reconstructs and identifies each individual particle with an optimized combination of
information from the various elements of the CMS detector. The energy of photons is directly obtained from the ECAL measurement, corrected for zero-suppression effects.  The
energy of electrons is determined from a combination of the electron momentum at the primary interaction vertex as determined by the tracker, the energy of the corresponding ECAL
cluster, and the energy sum of all bremsstrahlung photons spatially compatible with originating from the electron track.
The momentum of muons is obtained from the curvature of the corresponding track as determined by the tracker and muon system.  The energy of charged hadrons is determined from a combination of their momentum measured in the tracker and the matching ECAL and HCAL energy deposits,
corrected for zero-suppression effects and for the response function of the calorimeters to hadronic showers. Finally, the energy of neutral hadrons is obtained from the
corresponding corrected ECAL and HCAL energies.

Jets are reconstructed using the Cambridge--Aachen (CA) \cite{CAcambridge,CAaachen} algorithm with a distance parameter
of 0.8 (CA8 jets) as implemented by FastJet 3.0.4 \cite{fastjet1,fastjet2} to
cluster PF candidates into jets.
This algorithm clusters constituents (the reconstructed PF candidates in each event) to form jets based only on the angular distance between them.
The CA algorithm has been shown to have higher efficiency for distinguishing jet substructure~\cite{catop_cms} than competing jet clustering algorithms.

Jet momentum is determined as the vector sum of all particle momenta in the jet, and is found in the simulation to
be equal to the true momentum at hadron level within 5\% to 10\% over the full $\pt$
spectrum and detector acceptance.

The jet energy resolution amounts typically to 15\% at 10\GeV, 8\% at 100\GeV, and
4\% at 1\TeV, to be compared to about 40\%, 12\%, and 5\% obtained jet clustering is based on calorimeter information only rather than on PF candidates.

The jet energy in simulation is corrected using measurements derived from data.
The jet energy corrections for the CA jets are derived from the anti-\kt (AK) jet clustering algorithm ~\cite{Cacciari:2008gp} with a distance parameter value of 0.7.
The AK jet energy corrections have been shown to be applicable to CA jets~\cite{Chatrchyan:2013lca}.
The uncertainty associated with this procedure is noted in Section \ref{sec:systematics}.

We use the charged hadron subtraction method~\cite{CMS-PAS-JME-14-001} to remove charged hadrons that
originate from a non-leading vertex prior to the application of the jet clustering algorithm, where the leading vertex is defined as the vertex with the largest sum of $\pt^2$.
After this procedure, the neutral component of pileup (inelastic proton-proton pair interactions in the same bunch crossing) is subtracted using an area based method~\cite{Cacciari:2007fd}.

\subsection{Signal modeling}

The signal samples are generated at leading order with the CompHEP \cite{CompHEP} package and then scaled to next-to-leading order
using a factor of 1.2 \cite{kfactor}.  We generate signal samples using three coupling hypotheses (see Eq. (\ref{eqn:Lag})):

\begin{itemize}
\item \textbf{\boldmath{$\wprr$}} --- purely right-handed $\wpr$ boson where $a_{\mathrm{q}_i\mathrm{q}_j}^{\mathrm{L}}$=0 and $a_{\mathrm{q}_i\mathrm{q}_j}^{\mathrm{R}}$=1;
\item \textbf{\boldmath{$\wprl$}} --- purely left-handed $\wpr$ boson where $a_{\mathrm{q}_i\mathrm{q}_j}^{\mathrm{L}}$=1 and $a_{\mathrm{q}_i\mathrm{q}_j}^{\mathrm{R}}$=0;
\item \textbf{\boldmath{$\wprlr$}} --- mixed-coupling $\wpr$ boson where $a_{\mathrm{q}_i\mathrm{q}_j}^{\mathrm{L}}$=1 and $a_{\mathrm{q}_i\mathrm{q}_j}^{\mathrm{R}}$=1.
\end{itemize}

The $\wprr$ width varies from 44 to $91\GeV$ for the mass range considered in this analysis.
The generation of the left-handed and mixed-coupling samples takes into account the interference with the SM $s$-channel single top quark production.

\subsection{Combined CMS t-tagging algorithm}
\label{sec:toptagging}
When the W boson decays to hadrons, the top quark can be detected as three jets.
The high boost of the top quark from a $\wpr$ boson decay causes the three jets to merge into one large jet with a distinct substructure.
The CMS t-tagging algorithm~\cite{JME13007} discriminates signal from background by using this characteristic substructure.
The algorithm reclusters the CA jet until it finds anywhere from 1 to 4 subjets
In this process, particles with low $\pt$ or at a large angular distance from the jet centre are omitted.
The t-tagging algorithm is based on the following selection:

\begin{itemize}
\item \textbf{Jet mass   \boldmath{$ 140 < M_{\text{jet}} < 250$}\GeV} --- The mass of the CA jet is required to be consistent with the top quark mass.
\item \textbf{Number of subjets  \boldmath{$ N_{\text{subjets}} > 2$}} --- The number of subjets found by the algorithm must be at least 3.
\item \textbf{Minimum pairwise mass \boldmath{$ M_{\text{min}} > 50$}\GeV}  --- The three highest $\pt$ subjets are
taken pairwise, and the pair with the lowest invariant mass is calculated ($M_{\text{min}}$).
The value of $M_{\text{min}}$ is required to be greater than 50\GeV for consistency with the W boson mass.
\end{itemize}

In addition to these requirements, the N-subjettiness algorithm is used for t-jet identification \cite{Thaler:2011gf}.  This algorithm defines the $\tau_N$ variables,
which describe the consistency between the jet energy and the number of assumed subjets, $N$:

\begin{equation}
	\tau_{N} = \frac{1}{d}\sum_{i}p_{\mathrm{T}_i}\min\{\Delta R_{1,i},\Delta R_{2,i},\ldots,\Delta R_{N,i} \},
\end{equation}
where $\Delta R_{\mathrm{J},i}$ is the angular distance ($\Delta R = \sqrt{\smash[b]{(\Delta\eta)^2+(\Delta\phi)^2}}$) measured between the subjet candidate (J) axis and a specific constituent particle ($i$),
and $d$ is the normalization factor:
\begin{equation}
	d = \sum_{i} p_{\mathrm{T}_i} R,
\end{equation}
where $R$ is the characteristic distance parameter used by the jet clustering algorithm.
A jet with energy consistent with $N$ subjets will typically have a low $\tau_{N}$ variable.  A $\PQt$ jet should be more consistent with three subjets than two
(when compared to jets originating from a gluon or a light quark), therefore the ratio
of $\tau_3$ and $\tau_2$ allows top jets to be distinguished from
multijet events from QCD processes (labeled in the following as QCD background).  We select events with $\tau_3/\tau_2 < 0.55$.

We apply the combined secondary vertex (CSV) $\PQb$-tagging algorithm to
all of the subjets found by the t-tagging algorithm.  We require the maximum CSV discriminator value
to pass CSV b tagging at the medium operating point ($\mathrm{SJ}_{\text{CSVMAX}} \geq 0.679$)~\cite{CMS-PAS-BTV-13-001}.
The CMS t tagger with the addition of N-subjettiness and subjet b tagging is referred to as the combined CMS t tagger.

Substructure variables in the signal region exhibit known differences between
data and simulation that affect the t-tagging efficiency. We derive
a scale factor that is the ratio of the t-tagging efficiency measured in
data to simulation~\cite{JME13007}, using hadronic top quark decays from a control region that consists of an almost pure sample of
semileptonic $\ttbar$ events. This ratio, which is measured to be
$1.04 \pm 0.13$, is applied as a correction factor to the signal samples that are used in this analysis.

\subsection{Identification of b jets}
\label{sec:sjbtag}

\label{sec:btagging}

To identify the b quark daughter of the $\wpr$ boson, we start with the $\PQb$-candidate jet,
which is the highest $\pt$ jet that is hemispherically separated from the top-tagged jet.
We apply the CSV algorithm used to identify b jets to this $\PQb$-candidate jet. The medium operating
point is used, which has a light-flavour jet misidentification probability
of 1\% for an efficiency around 70\%. A scale factor is applied to correct for differences
in $\PQb$-tagging efficiency between data and simulation~\cite{CMS-PAS-BTV-13-001}. The uncertainties in
this scale factor are described in Section~\ref{sec:systematics}.  Backgrounds from
SM $\ttbar$ production are reduced by requiring the invariant
mass of the $\PQb$-candidate CA jet to be below $70\GeV$.

\subsection{Reconstruction of \texorpdfstring{$\wpr$}{W'} mass}
\label{sec:full-selection}
We select candidate $\mathrm{\wpr \to tb}$ events by using the following criteria, which are applied to the two leading jets:
\begin{itemize}
\item One jet with $\pt > 450\GeV$ identified with the combined CMS t-tagging algorithm;
\item One jet with $\pt > 370\GeV$ with a CSV b tag at the medium operating point and mass $<$70\GeV;
\item The two jets are in opposite hemispheres ($\abs{\Delta \phi} > \pi/2$);
\item The difference in rapidity between the two jets ($\abs{\Delta y}$) is less than 1.6.
\end{itemize}
The number of events after each successive selection remaining in data, SM $\ttbar$ production, $s$-channel single top,
and right-handed $\wpr$ boson signal simulations is shown in Table \ref{table:Cutflow}.

\section{Event samples}
\label{sec:datasampleAndSelection}
The data used for this analysis correspond to an integrated luminosity of $19.7\fbinv$ of pp collisions
provided by the LHC at a centre-of-mass energy of $8\TeV$.  We select events online using a trigger algorithm that requires the scalar $\pt$ sum of reconstructed jets in the detector to be
$>$750\GeV. The trigger is nearly 100\% efficient for events selected in the offline analysis. The small trigger inefficiency is
measured from data and applied to the simulated events.

\label{sec:preselection}
\label{sec:reconstruction}

\begin{table}[htb]
\centering
\topcaption{Numbers of observed and expected events at successive stages of the event selection.
The expected numbers are scaled to an integrated luminosity of 19.7\fbinv.  Statistical uncertainties in the event yields are quoted.
The QCD background contribution is only reported for the final selection.  The first row implies a 150\GeV \pt selection on the jets, the trigger selection,
and the requirement that the two leading jets be in opposite hemispheres.
The row labeled ``$\pt$'' represents the transverse momentum requirements placed on the two leading jets. The signal events, shown for several
values of the $\wpr$ boson mass, are normalized to the theoretical cross section.  The $s$-channel single top contribution ($\mathrm{ST}_{s}$) is given as well as it
is used when deriving the shape of the signal distribution when calculating the generalized coupling limits.}
\newcolumntype{x}{D{,}{\,\pm\,}{-1}}
\resizebox{\textwidth}{!}{
\begin{tabular}{l|rcxxxxxx}
\hline
Selection & Data & QCD & \multicolumn{1}{c}{$\ttbar$} &\multicolumn{1}{c}{$\mathrm{ST}_s$}& \multicolumn{1}{c}{\begin{tabular}{@{}c@{}} $M_{\wprr}$\\ 1.90$\TeV$\end{tabular}} & \multicolumn{1}{c}{ \begin{tabular}{@{}c@{}} $M_{\wprr}$\\ 2.10$\TeV$\end{tabular} }  & \multicolumn{1}{c}{\begin{tabular}{@{}c@{}} $M_{\wprl}$\\ 1.90$\TeV$\end{tabular}} & \multicolumn{1}{c}{\begin{tabular}{@{}c@{}} $M_{\wprl}$\\ 2.10$\TeV$\end{tabular} }  \\
\hline
2 jets & 13854873 & --- & 12190,27  & 283,8.6 & 806,1 & 401,0.7 & 796,2 & 430,2 \\
$\pt$ & 4305244 & --- & 4720,18  & 130,6.5  & 739,1 & 372,0.7 & 703,2 & 364,2 \\
$\abs{\Delta \mathrm{y}}$ & 3376771 & --- & 4220,17  &  121,6.3 & 553,1 & 268,0.6 & 531,2 & 268,1 \\
$M_{\PQt}$ & 992949 & --- & 3220,14  &  64,4.5 & 429,1 & 209,0.5 & 414,2 & 205,1  \\
$N_{\text{subjets}}$ & 557489 & --- & 2740,13  & 48,3.9 & 340,0.9 & 163,0.5 & 312,2 & 152,1  \\
$M_{\text{min}}$  & 318520 & --- & 2510,13  &  42,3.7 & 304,0.9 & 143,0.4 & 274,2 & 130,0.9 \\
$\mathrm{SJ}_{\mathrm{CSVMAX}}$  & 50642 & --- & 1690,10  & 23,2.6  & 170,0.6 & 76,0.3 & 138,1 & 63,0.6  \\
$\tau_3/\tau_2$ & 7200 & --- & 1024,8  &  11,1.8  & 88,0.5 & 38,0.2 & 58,0.7 & 27,0.4 \\
$M_{\PQb}$ & 4463 & --- & 178,4  &  8,1.6 & 68,0.4 & 29,0.2 & 44,0.6 & 20,0.3  \\
CSV & 277 & 248${\,\pm\,}$4 & 37,1  &  2,0.71  & 16,0.2 & 6,0.1 & 10,0.3 & 4,0.2 \\
\hline
\end{tabular}}
\label{table:Cutflow}
\end{table}

\section{Background modeling}
\label{sec:backgroundEstimation}
The primary sources of background are SM QCD multijet and $\ttbar$
production.  These backgrounds dominate because of QCD background events after selecting an all-jet final state is selected
and the large contribution from $\ttbar$ production that remains after t-jet discrimination criteria are applied.

The principal background in this analysis is QCD multijet production, and is estimated using a data-driven technique to extract both the shape and the normalization.
The average $\PQb$-tagging rate, measured from events with an enhanced QCD multijet component and a negligible signal contamination component, is
used to estimate the QCD multijet contribution in the signal region.  We account for the background contribution from SM $\ttbar$ production
when measuring the $\PQb$-tagging rate.  In addition to the region used to extract the average $\PQb$-tagging rate, we define two test regions to check the QCD background prediction, both of which have small
$\ttbar$ background and possible signal contamination.

The shape of the $\mtb$
distribution for $\ttbar$ production is estimated from Monte Carlo (MC) simulation,
and the yield is measured from data using a control sample enriched in $\ttbar$ events.

\subsection{QCD background estimate}
\label{sec:qcdBackgroundEstimationProcedure}

A control sample
is obtained by inverting the $N_{\text{subjets}}$ selection criteria used to
identify t jets:
\begin{eqnarray}
	140  <  M_{\text{jet}}  <  250\GeV; \\
	N_{\text{subjets}}  \leq  2;\\
	\mathrm{SJ}_{\text{CSVMAX}} \geq 0.679.
\end{eqnarray}

We apply the b tagging criteria to the $\PQb$-candidate jet in the event to measure the average $\PQb$-tagging rate for QCD jets.
We assume that this rate is the same for QCD jets in the signal region.
We include the subjet CSV discriminant in this control sample in order to ensure similar parton
flavour distributions in the signal and control regions.
Because of the similar parton flavour distributions, and the fact that this region has a low $\ttbar$ and signal contamination component, this control sample is an ideal selection to extract the average $\PQb$-tagging rate.
The average $\PQb$-tagging rate is parameterized as a function of the $\pt$ of the $\PQb$-candidate jets (which pass all requirements except the b tag)
in three $\abs{\eta}$ regions:

\begin{itemize}
	\item \text{Low } $(0.0 < \abs{\eta} \leq 0.5)$;
	\item \text{Transition } $(0.5 < \abs{\eta} \leq 1.15)$;
	\item \text{High } $(1.15 < \abs{\eta} \leq 2.4)$.
\end{itemize}

Events in the signal region that do not have b tagging applied are then weighted by this average $\PQb$-tagging rate to
estimate the QCD background contribution in the final selection.
Events in the signal region before b tagging is applied are largely from QCD background, but the small $\ttbar$
background component (less than 1\%) is subtracted when deriving the QCD background contribution to avoid double counting.

\label{sec:tagrateparameterization}

\begin{figure}[htbp]
\centering
\includegraphics[width=0.65\textwidth]{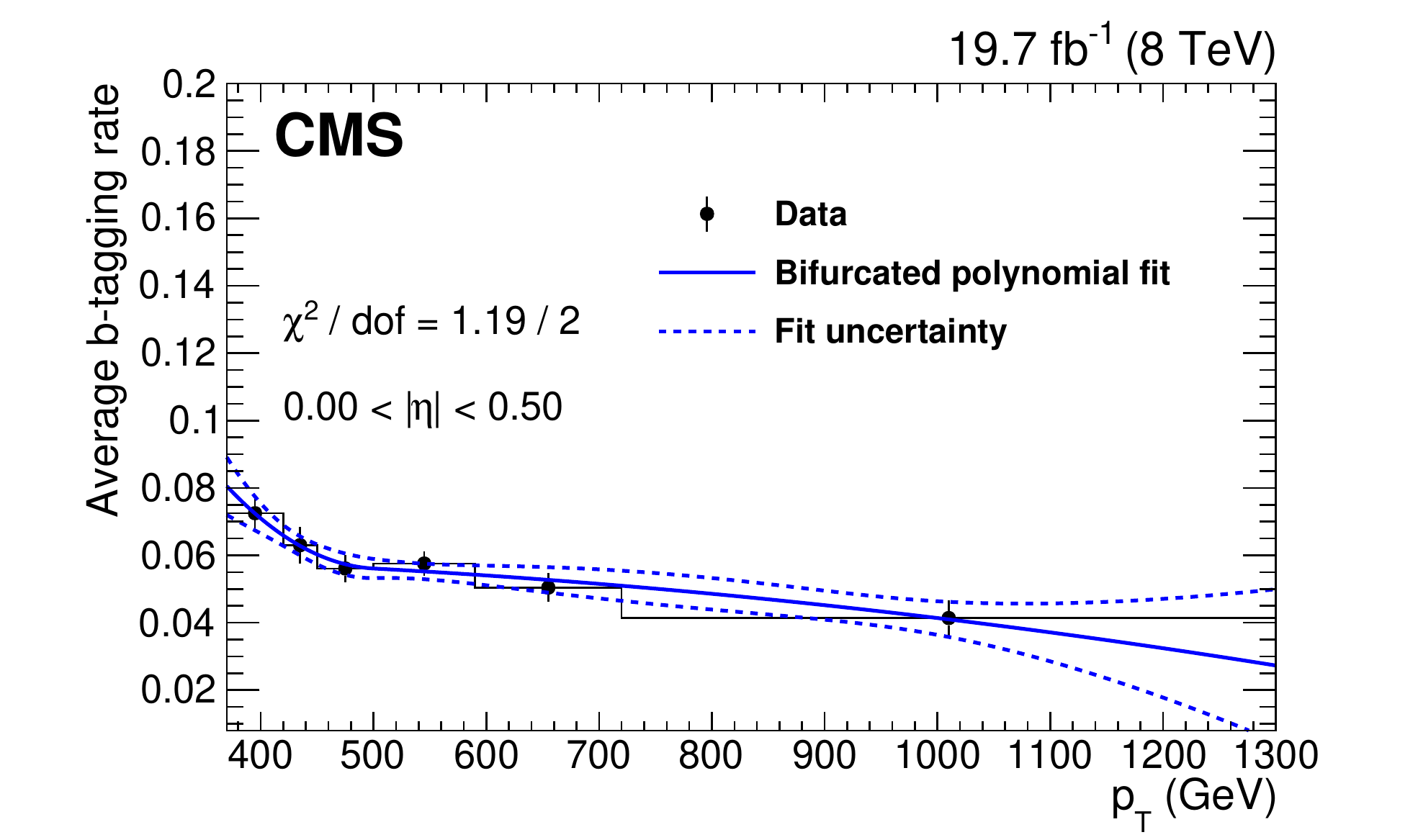}\\
\includegraphics[width=0.65\textwidth]{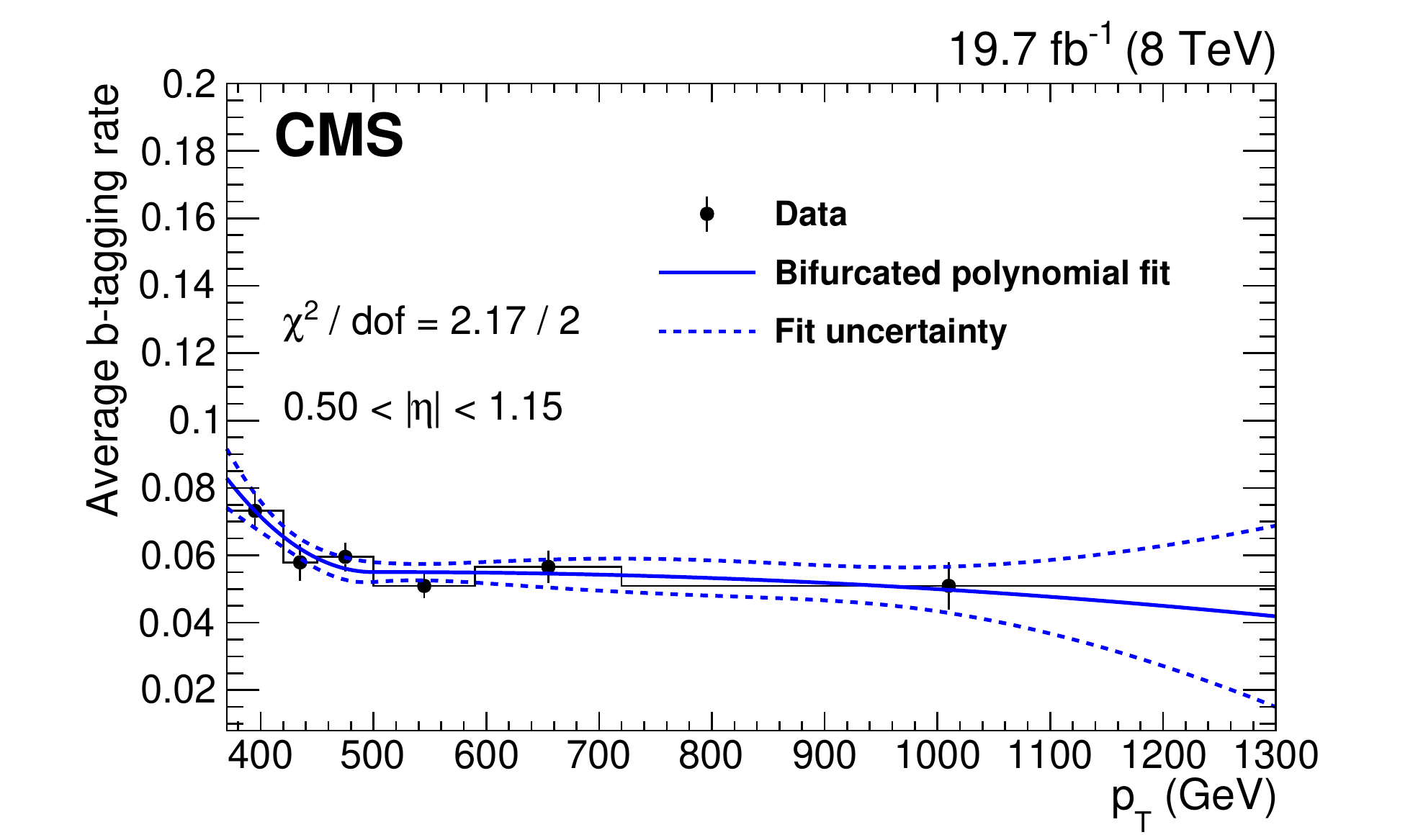}\\
\includegraphics[width=0.65\textwidth]{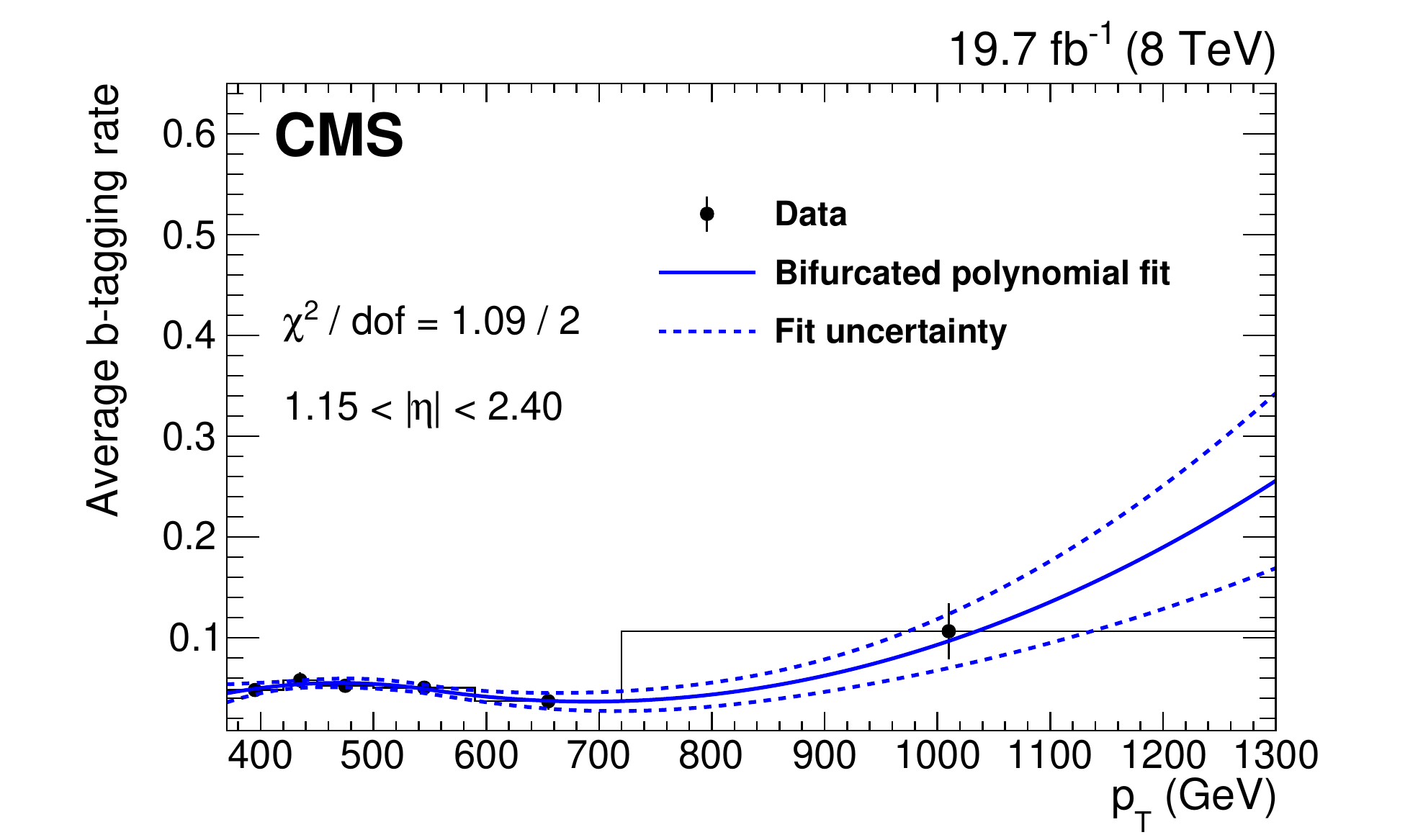}
\caption{
The average $\PQb$-tagging rate for QCD jets parameterized as a function of the $\pt$ of the $\PQb$-jet candidate from the low (top), transition (middle), and high (bottom) $\abs{\eta}$ regions.
  The measured average $\PQb$-tagging rate is represented by the data points, the polynomial fit is shown as a solid line, and the propagated uncertainties from the fit are shown as the dashed lines.
The horizontal lines on the data points indicate the bin widths.}
\label{figs:tagrateetafit}
\end{figure}

We use a bifurcated polynomial to fit the average $\PQb$-tagging rate in each of these $\abs{\eta}$ ranges.
This fitting function, which provides a satisfactory description of the data, is defined as follows:

\begin{eqnarray}
f(x) =
\begin{cases}
p_0+p_1x+p_2(x-a)^2, & \text{if}~x < a \\
p_0+p_1x+p_3(x-a)^2, & \text{if}~x \geq a
\end{cases}
\end{eqnarray}

Here, the parameters $p_0$ to $p_3$ are the polynomial coefficients, and $x$ is the $\pt$ of the $\PQb$-candidate jet.
The parameter $a$ is the bifurcation point, and is optimized for each region in $\abs{\eta}$.
It is chosen to be 500, 500, and 550\GeV for the low, transition, and high $\abs{\eta}$ regions, respectively.
The parameterization of the average b-tagging rate helps to constrain the known $\pt$ and $\abs{\eta}$ kinematic correlation inherent in b tagging, which is due to detector geometry and tracking resolution.

The uncertainty in the average $\PQb$-tagging rate is extracted using the full covariance matrix obtained from the output of the fitting
algorithm.  Additionally, we assign a systematic uncertainty to cover the choice of the fit function
(see Section \ref{sec:systematics}) based on several alternative functional forms (such as second degree polynomial or exponential functions).
Fig. \ref{figs:tagrateetafit} shows the measured average $\PQb$-tagging rate and associated uncertainty bands.

\subsection{The \texorpdfstring{$\ttbar$}{t-tbar} background estimate}
\label{sec:ttbarsideband}
We obtain the normalization of the background contribution from SM $\ttbar$
production using a control sample in data. A sideband region is defined (CR3) by
inverting the $\PQb$-candidate mass requirement (see Section~\ref{sec:btagging}) in the signal region.
This region has an enhanced fraction of $\ttbar$ events and is
statistically independent from all other sidebands in the analysis.

We compare the sum of the QCD background estimate from data and
the SM $\ttbar$ contribution obtained from MC simulation to the
observed yield in data. The $\ttbar$ and single top quark production samples used
in this analysis are generated using the POWHEG~\cite{Alioli:2010xd,Frixione:2007vw,Nason:2004rx} 1.0 event generator and are normalized to the SM
expectation using next-to-next-to leading-order
cross sections~\cite{Czakon:2013goa,Kidonakis:2012db}.

For the QCD background determination in this sideband region, we use the procedure outlined in Section~\ref{sec:qcdBackgroundEstimationProcedure}.
However, we invert the $\PQb$-candidate mass requirement when extracting the average $\PQb$-tagging rate in order to account for potential correlations between
the $\PQb$-candidate mass and the $\PQb$-tagging rate.

\begin{figure}[htbp]
\centering
\includegraphics[width=0.6\textwidth]{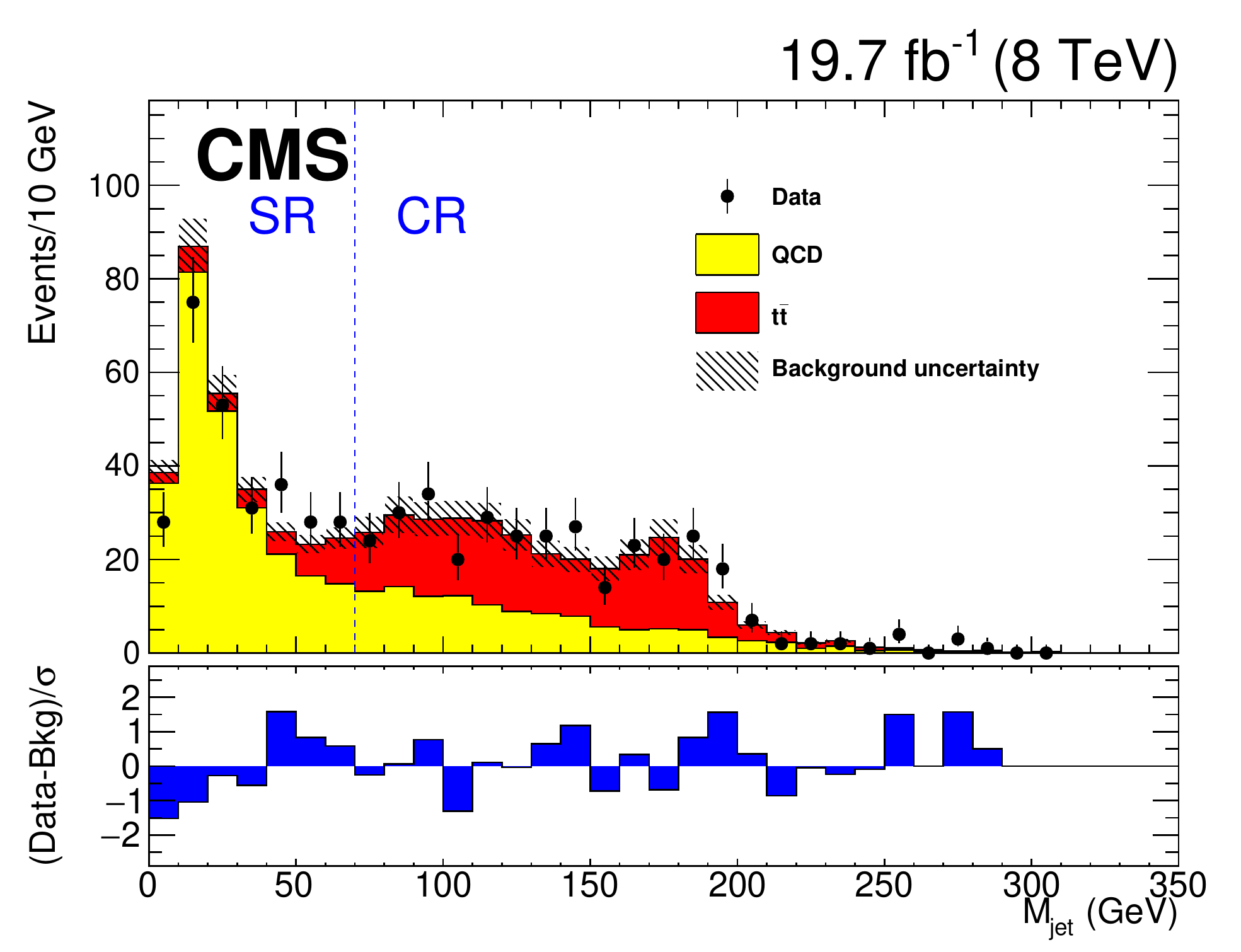}
\caption{Distribution of the mass of the $\PQb$-jet candidate after the template fit to constrain the QCD multijet and $\ttbar$ backgrounds has been applied. A control region with inverted
selection criteria on the mass of the $\PQb$-jet candidate is used for the fit, and is shown as the area to the right of the vertical dashed line.  The hatched region represents the background uncertainty, obtained by adding the QCD multijet uncertainty and the uncertainty on the $\ttbar$ normalization from the output of the fit in quadrature.
 The bottom plot shows the pull ((data-background)/$\sigma$) between the data and the background estimate distributions.}
\label{figs:ttbarfit}
\end{figure}

We perform a binned maximum likelihood fit to the invariant mass distribution
of the $\PQb$-candidate jets, using the shape of the $\ttbar$ background MC prediction as
one template, and the QCD background prediction from data as the other.
The normalization of the QCD background template is allowed to vary within its systematic uncertainty envelope,
whereas the normalization of the $\ttbar$ template is left unconstrained. The result
of the fit is shown in Fig.~\ref{figs:ttbarfit}.

The contamination from $\ttbar$ events must be taken into account when obtaining the QCD background template in this sideband from data.
The fit described above independently varies the QCD and $\ttbar$ templates,
however the component of the QCD background from $\ttbar$ events introduces a small anticorrelation between the two templates.
We account for this by first fitting the QCD background before the
$\ttbar$ component is subtracted and correcting the resulting $\ttbar$ normalization by the factor 1+C/S, where C/S is the ratio of
the nominal number of $\ttbar$ events in the control region (C) to that in the signal region (S).  After applying this correction, the $\ttbar$
normalization obtained from the fit is independent of the fraction subtracted from the QCD background template.
Following this procedure we find that the SM $\ttbar$ production rate in the signal region must be scaled by a factor of $1.23\pm0.24$.

\label{sec:ttptrw}
In order to correct for known differences in the top quark $\pt$ spectrum between data and MC simulation of
SM $\ttbar$ production~\cite{CMS-PAS-TOP-12-027}, we reweight the MC events using the generator
level $\pt$ of the top quark and top anti-quark.  Although this procedure was not designed for the kinematic
range in our analysis, we still use it as the change to the normalization of the $\ttbar$ background template is
consistent with our measurement of this extra normalization factor.

\subsection{Control region closure test}
\label{sec:secondsideband}
\begin{figure}[htbp]
\centering
\includegraphics[width=0.6\textwidth]{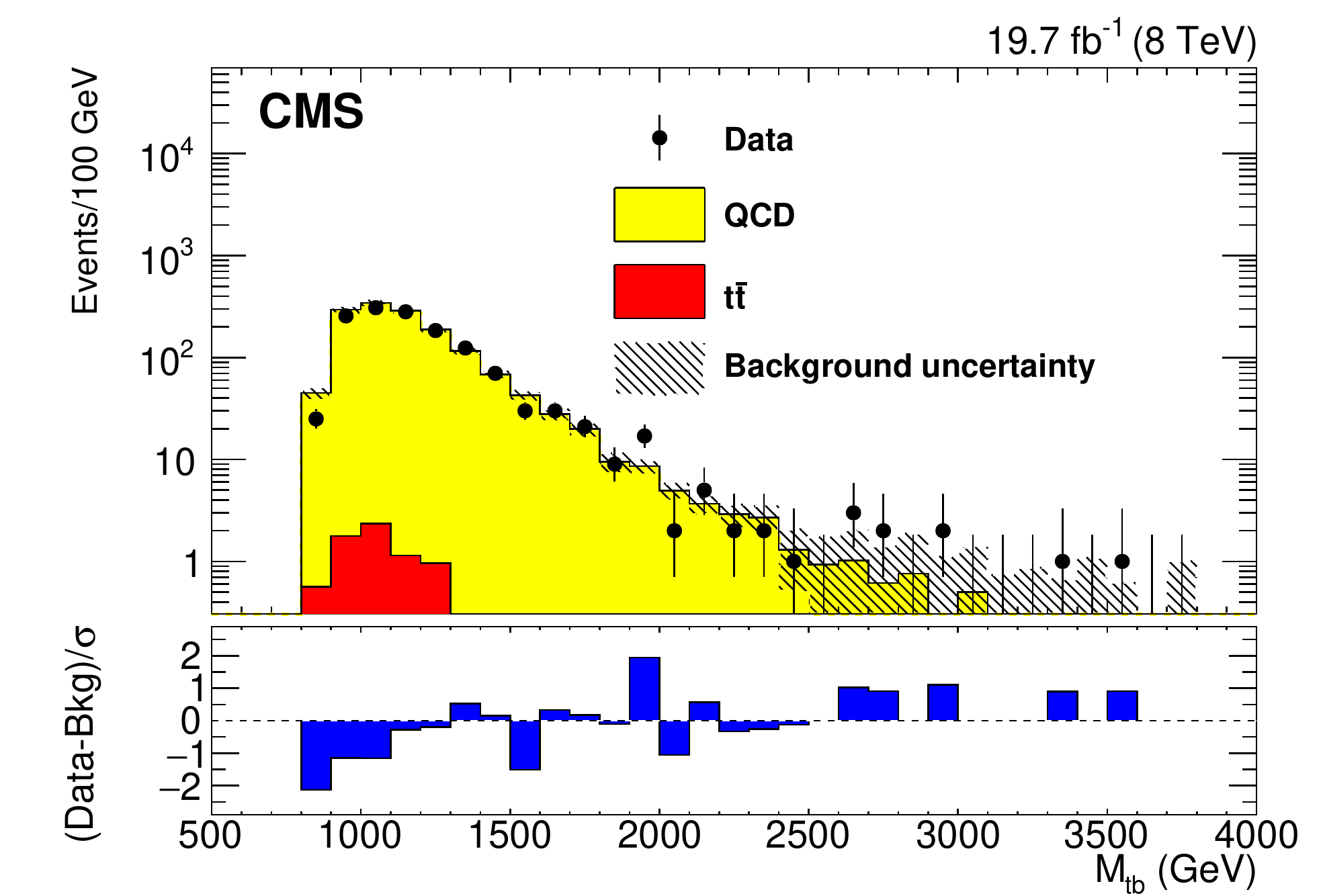}
\includegraphics[width=0.6\textwidth]{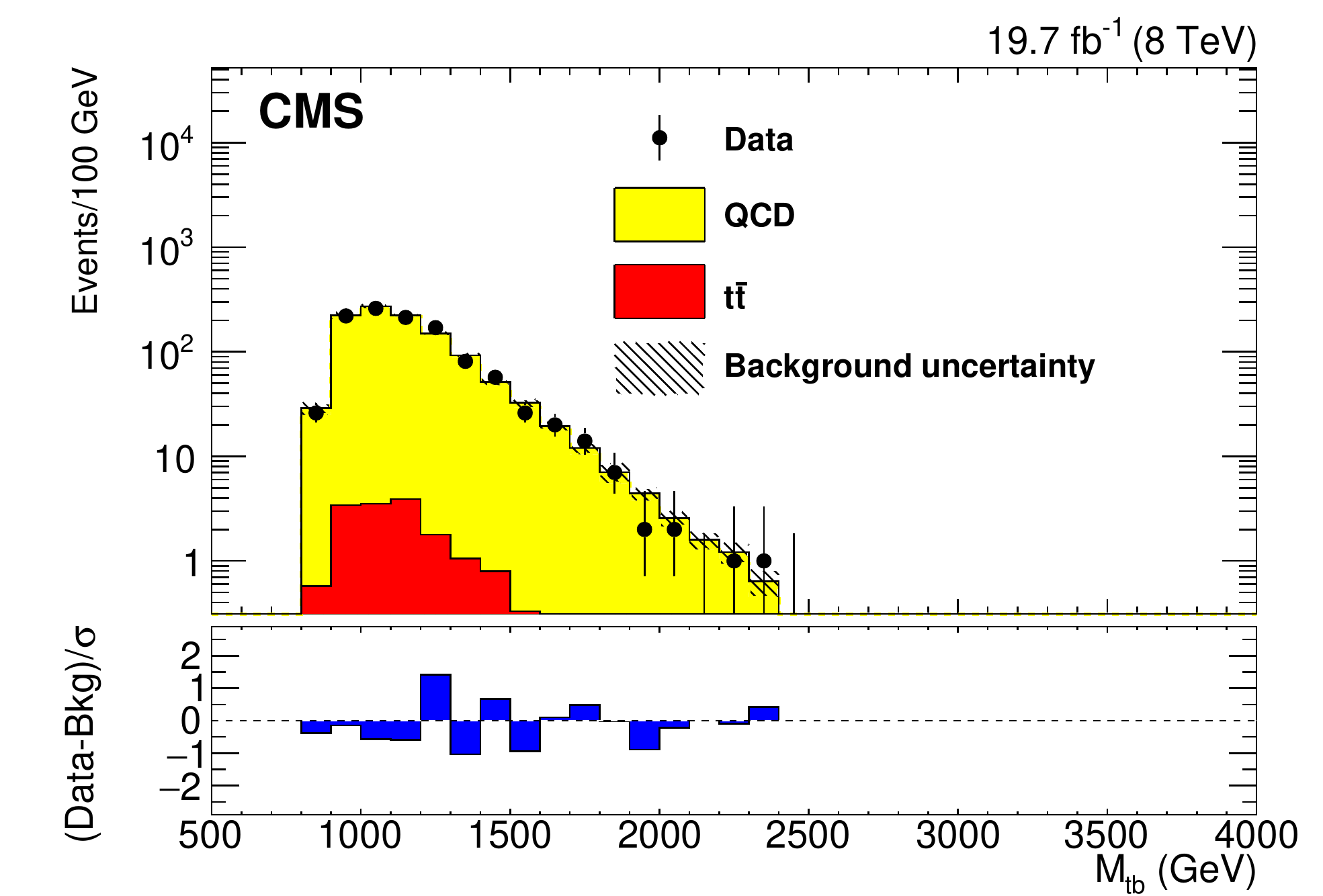}
\caption{Distributions of $\mtb$ shown for data, $\ttbar$, and QCD in the control regions CR1 (top) and CR2 (bottom) as defined in the text.
The hatched region shows the background uncertainty, obtained by adding the statistical and systematic uncertainties in quadrature.  The bottom plots show the pull ((data-background)/$\sigma$) between the data and the background estimate distributions.}
\label{figs:NewMtbSB3}
\end{figure}

To investigate the applicability of the QCD background estimation in data, we apply the average $\PQb$-tagging
rate to control regions of the t-tagging selection defined in Section \ref{sec:toptagging}.
First, we define a control region with inverted minimum pairwise mass and N-subjettiness selections (CR1).
This region is orthogonal to both the signal region and the control region that is used for the determination of the average $\PQb$-tagging rate.
The selection also has a very low yield of $\ttbar$ production, which allows for a precise measurement of the QCD background contribution.
The average $\PQb$-tagging rate used for this closure test is extracted from the same control region as the signal region, and is applied to events that are not b tagged.
This test shows good agreement as shown in Fig. \ref{figs:NewMtbSB3} (top).  Additionally, we define a control region with an inverted subjet $\PQb$-tagging selection (CR2).
This test also shows good agreement as seen in Fig. \ref{figs:NewMtbSB3} (bottom).

These control regions have low signal contamination.  For the 1.90$\TeV$~$\wprr$ sample, the signal contamination is less than 1\%.
These control regions are summarized in Table~\ref{table:CoRe}.

\begin{table}[htb]
\centering
\topcaption{Numbers of observed and expected events for each of the sidebands used for closure tests of the
QCD and $\ttbar$ background estimates.  A summary of the inverted selection for each control region is provided.}
\begin{tabular}{l|ccc}
\hline
Control Region & Data & QCD & $\ttbar$   \\
\hline
CR1: $M_{\text{min}}<50\GeV$; $\tau_3/\tau_2>0.55$  & 1100 & 1107 & 15\\
CR2: $\mathrm{SJ}_{\text{CSVMAX}}<0.679$ & 1376 & 1466 & 8 \\
CR3: $M_{\mathrm{b}}>70\GeV$ & 336 & 121 & 200  \\
\hline
\end{tabular}
\label{table:CoRe}
\end{table}

\section{Results}
\label{sec:stats}

\begin{figure}[htbp]
\centering
\includegraphics[width=0.7\textwidth]{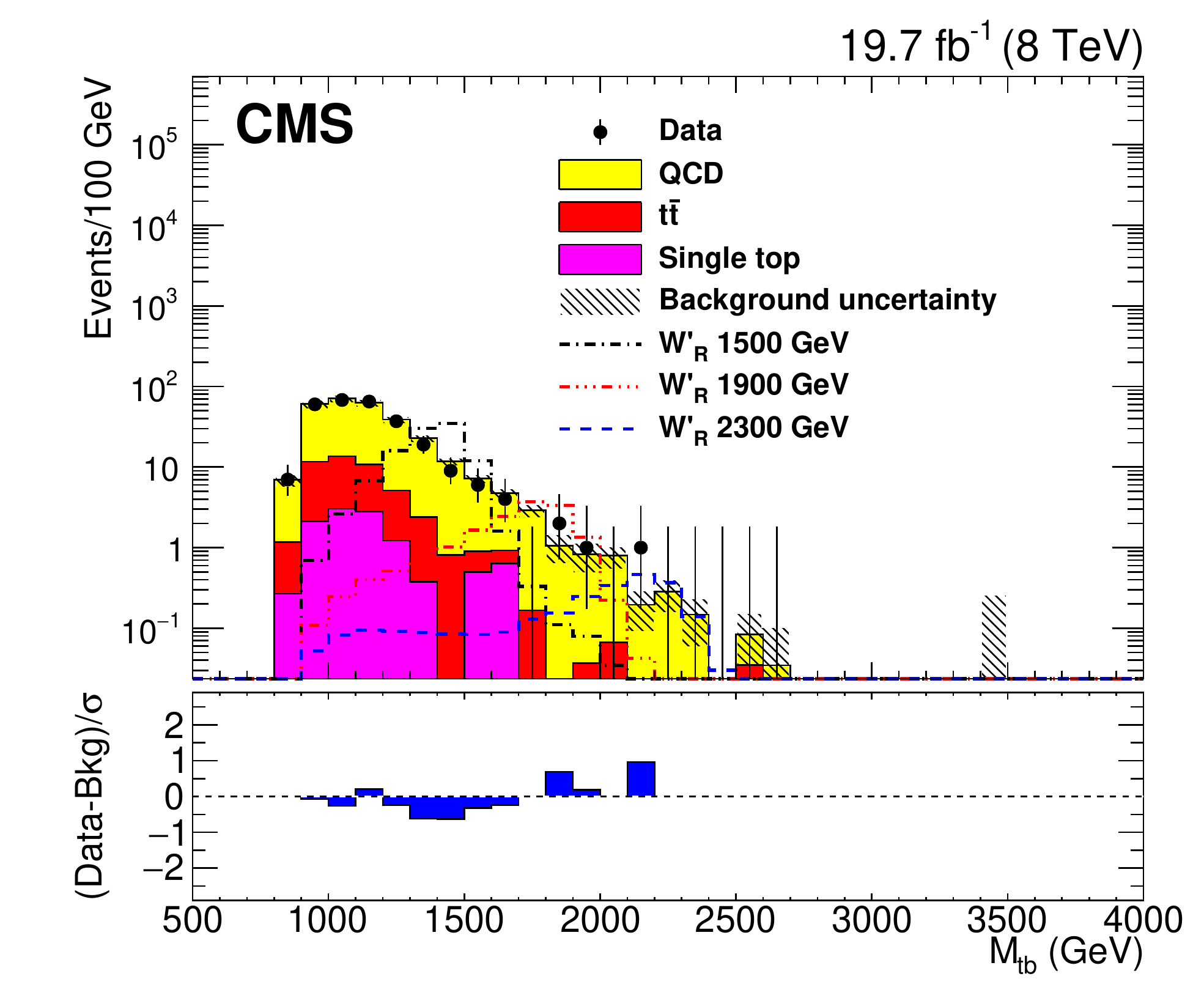}
\caption{
The distribution of $\mtb$ shown for data, $\ttbar$, QCD, single top, and several example signal $\wpr$ boson mass hypotheses.
The normalization for the $\wpr$ signal samples assumes the cross section from theory.
The distributions are shown after the application of all selection criteria.
The background contribution from single top quark production is not considered when setting limits.  The hatched region shows the background uncertainty, obtained by adding the statistical and systematic uncertainties in quadrature.
The bottom plot shows the pull ((data-background)/$\sigma$) between the data and the background estimate distributions.
}
\label{figs:MtbvsBkg1}
\end{figure}

After constraining the SM $\ttbar$ normalization using a particular control region (see Section~\ref{sec:ttbarsideband}),
and investigating the agreement of the data and the QCD background estimate
in two additional control regions (see Section~\ref{sec:secondsideband}),
the background estimate is used to predict the data distribution in the signal region, in the absence of a $\wpr$ boson contribution.  The results are shown in Fig. \ref{figs:MtbvsBkg1}.
Good agreement between data and expectation from SM processes are observed, with no signs of a new physics signal.

\subsection{Systematic uncertainties}
\label{sec:systematics}
We consider several sources of systematic uncertainty, corresponding to uncertainties in both the shape and normalization of the $\mtb$ distribution,
which are summarized in Table 2.

A 19\% normalization uncertainty is assigned to the estimate of the SM $\ttbar$ normalization (see Section~\ref{sec:ttbarsideband}).
The scale factor used to account for differences in the t-tagging efficiency
between MC and data has an uncertainty of 13\%~\cite{JME13007}. The uncertainty in the integrated luminosity is 2.6\%~\cite{CMS-PAS-LUM-13-001}.
Finally, the MC to data scale factor for the $\PQb$-tagging efficiency is evaluated using AK jets but is
applied to CA jets, which requires a systematic uncertainty of 2\% in addition to the $\PQb$-tagging scale factor uncertainty of around 8\%.

Several uncertainty sources contribute to the estimation of the average $\PQb$-tagging rate, which is used to evaluate the QCD background prediction.
We include a contribution due to the uncertainties obtained from the polynomial fit used to parameterize the $\PQb$-tagging rate, and investigate the effect of choosing
alternative functional forms in the fit. We also obtain an estimate of the uncertainty caused by the specific parameterization
of the average $\PQb$-tagging rate by studying an alternative parameterization using $\pt$, $\eta$, and $\mtb$.

We derive an uncertainty in the effect of $\pt$ reweighting on the shape of the $\ttbar$ background by taking the unweighted $\mtb$ distribution
as the +1$\sigma$ shape.

The uncertainty arising from the variation of the renormalization and factorization scales in $\ttbar$ production is evaluated from
MC samples generated with two times ($+1\sigma$) and one half ($-1\sigma$) the nominal renormalization and factorization scales.

The $\PQb$-tagging scale factor ${\pm}1\sigma$ values are applied to $\ttbar$ production and signal MC \cite{CMS-PAS-BTV-13-001}.
The nominal jet energy corrections created for use with AK jets are applied in the analysis.
The ${\pm}1\sigma$ uncertainties in jet energy scale and resolution arising from the application of AK jet energy scale corrections are also considered.
Additionally, a 3\% uncertainty is applied
to the jet energy scale to account for differences in CA and AK jets.
Finally, the uncertainty in the trigger turn-on efficiency that is applied to the MC samples is taken as one half of the trigger inefficiency.

Uncertainties arising from parton distribution functions are studied by varying the eigenvalues of the parton distribution functions that are used in the simulation.
Pileup modeling in simulation is corrected by comparing the number of pileup interactions to the mean number of interactions in data.
The uncertainty on this correction is studied by varying the minimum bias cross section.  These sources, along with the jet angular resolution variation, are negligible.

The two largest systematic effects arise from the uncertainties in the average b-tagging rate for the QCD
background (approximately a 6\% normalization effect) and in the top tagging scale factor for the signal (13\%).

\begin{table}[htbp]
\topcaption{Sources of uncertainty that affect the $\mtb$ distribution and the $\pm$1$\sigma$ variations that are used in the fit.}
\centering
\begin{tabular}{l|c}
\hline
Source & Variation\\
\hline
$\ttbar$ normalization & 19\% \\

$\PQt$-tagging scale factor & 13\% \\

Integrated luminosity & 2.6\% \\

anti-\kt to CA8 jets $\PQb$-tagging scale factor & 2\%  \\

Average $\PQb$-tagging rate fit & ${\pm}1\sigma (\pt, \eta)$  \\

Alternate functional forms for the average $\PQb$-tagging rate & ${\pm}1\sigma (\pt, \eta)$ \\

Parameterization choice for the average $\PQb$-tagging rate & ${\pm}1\sigma (\pt, \eta, \mtb)$ \\

$\pt$ reweighting & Shape ${\pm}1\sigma (p_{\mathrm{T}_{\mathrm{t}}},p_{\mathrm{T}_{\PAQt}})$ \\

Renormalization and factorization scales & 2$Q^2$ and 0.5$Q^2$ \\

$\PQb$-tagging scale factor & ${\pm}1\sigma (\pt)$ \\

Jet energy scale & ${\pm}1\sigma (\pt, \eta)$ \\

Jet energy resolution & ${\pm}1\sigma (\eta)$ \\

Trigger efficiency & ${\pm}1\sigma (p_{\mathrm{T1}} + p_{\mathrm{T2}})$\\
\hline
\end{tabular}
\label{table:shapeunc}
\end{table}

\subsection{Cross section limits}

To set limits on the production cross section of
the $\wprr$ boson model described in Eq. (\ref{eqn:Lag}), we compare,
for each bin in the $\mtb$ distribution, the numbers of observed and expected events.  The small background contribution from single top quark production is not considered when setting limits.  The following expression is used to
compute the expected contribution from $\wprr$ boson production:

\begin{eqnarray}
N_{\textrm{expected}} = \sigma_{\wprr} \, \mathcal{B}_{\mathrm{\wprr \rightarrow tb;W \to hadrons}} \: \varepsilon   \int L\, \rd t,
\end{eqnarray}
where $\sigma_{\wprr}$ is the $\wprr$ cross section, $\mathcal{B}_{\mathrm{\wprr \rightarrow \mathrm{tb;W \to hadrons}}}$ is the branching fraction
$\mathrm{\wprr \to tb}$ with the W boson decay constrained to the hadronic branching fraction, $\varepsilon$ is the signal
efficiency, and $\int L\, \rd t$ is the integrated luminosity of the data set.
We perform a binned maximum likelihood fit to compare the $\mtb$ distribution from data with the $\wprr$ boson signal
hypothesis, summed together with the SM distribution obtained from the background estimation procedure described in Section~\ref{sec:backgroundEstimation}.

A Poisson model is used for each bin of the $\mtb$ distribution.  The mean of the Poisson distribution for each bin is taken to be:
\begin{eqnarray}
\mu_{i} = \sum_{k} \beta_{k} \, T_{k,i},
\end{eqnarray}
where $k$ includes both the signal and background models, $\beta_{k}$ is the Poisson mean for process $k$, and $T_{k,i}$ represents the fraction of events expected for each process $k$ in bin $i$.

The likelihood function is then:
\begin{eqnarray}
L(\beta_{k}) = \prod^{N_{\mathrm{bins}}}_{i} \frac{\mu_{i}^{N^{\mathrm{data}}_{i}} \, \re^{-\mu_{i}}}{(N^{\mathrm{data}}_{i})!},
\end{eqnarray}
\label{sec:Theta}
where $N^{\mathrm{data}}_{i}$ is the number of events in data for bin $i$.

\begin{figure}[htb]
\centering
\includegraphics[width=0.7\textwidth]{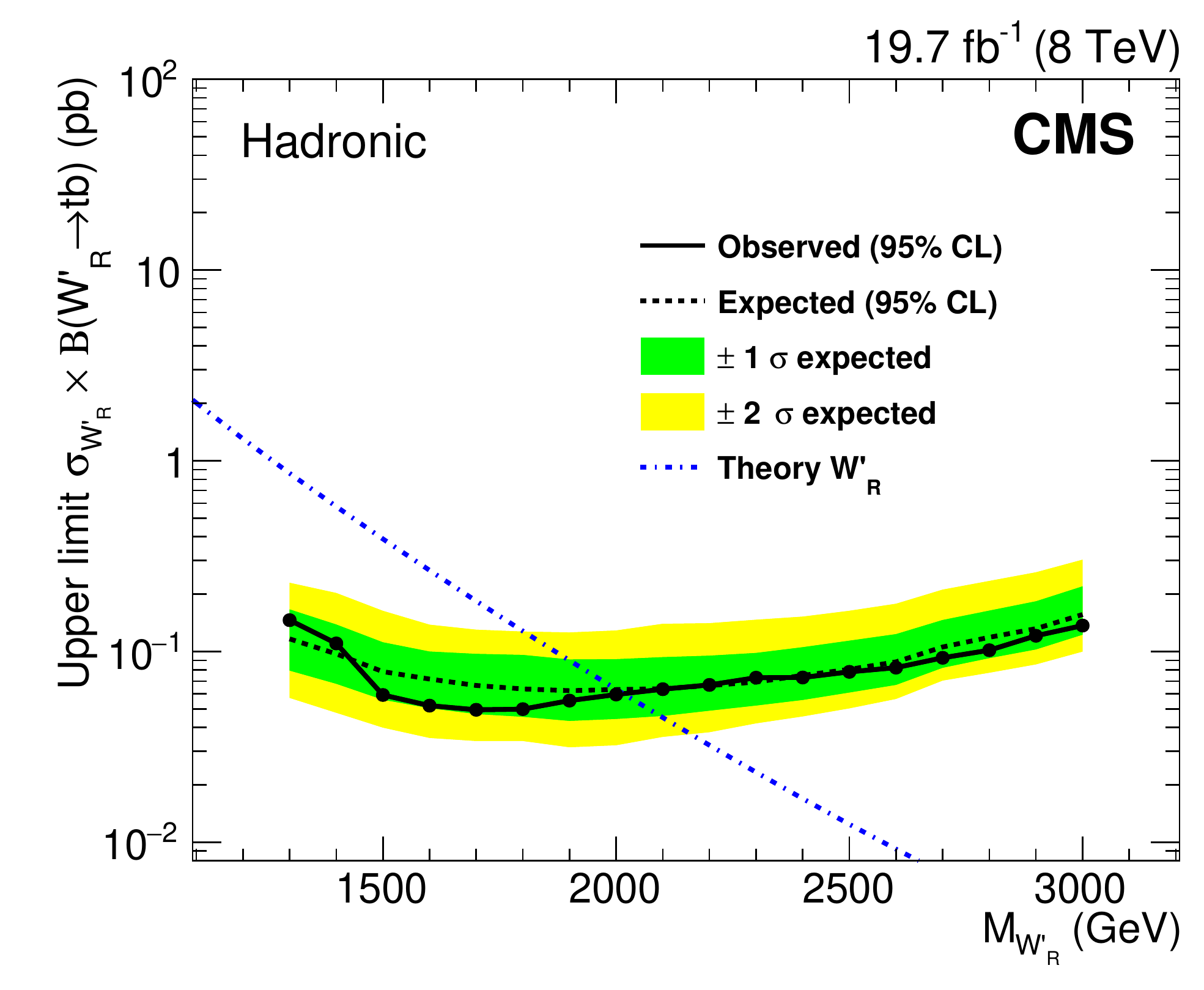}
\caption{The $\wprr$ boson 95\% CL production cross section times $\mathcal{B}_{\mathrm{\wprr \rightarrow tb}}$ limits for the hadronic channel.  The observed (solid) and expected (dashed) limits, as well as the $\wprr$ boson theoretical cross section (dot-dashed) are shown.
As indicated in the legend, the shaded regions about the expected limits represent 1 and 2$\sigma$ bands.}
\label{figs:thetalimit}
\end{figure}

Using a Bayesian approach with a flat prior for the signal cross section, we obtain 95\% confidence level (CL) upper limits on the production cross section of $\wprr$.
Pseudo-experiments are used to derive the ${\pm}1 \sigma$ deviations in the expected limit.
The systematic uncertainties described above are accounted for by nuisance parameters and the posterior probability is refitted for each pseudo-experiment.
The cross section upper limits are shown in Fig. \ref{figs:thetalimit}.  We exclude a $\wprr$ boson with a mass less than $2.02\TeV$ at 95\% CL .

We combine the results from the hadronic and leptonic
$\wpr$ decay modes to enhance the sensitivity of
the analysis to $\wpr$ production and the measurement of the coupling strengths of the $\wpr$ boson to quarks.
The analysis of the leptonic channel excludes a $\wpr$ mass below 2.05$\TeV$, and is described in Ref.~\cite{Chatrchyan:2014koa}.
The all hadronic and leptonic channels have similar sensitivity in the high $\wpr$ mass regime, which leads to a large increase in sensitivity in the combined result.

The hadronic channel probes $\wpr$ signal generated from a
mass of 1.3 to $3.1\TeV$ because of the sensitivity of the boosted top jet tagging techniques, whereas the leptonic channel probes $\wpr$ masses from 0.8 to $3.0\TeV$ because of the higher sensitivity at low $\wpr$ mass.  Therefore, the region of
combined sensitivity ranges from a $\wpr$ mass of 1.3 to $3.0\TeV$.  Below this region, only the leptonic channel limits are quoted.

There are points within the region of combined sensitivity where the signal sample exists for the leptonic channel but not
for the hadronic channel.  These
intermediate mass points are reproduced using \textsc{RooFit}~\cite{Verkerke:2003ir} template morphing to interpolate the shape of the $\mtb$ spectrum.
The generator level selection on the $\pt$ of the b quark for the left-handed and mixed-coupling $\wpr$ samples is taken into account by interpolating the
selection efficiency for the intermediate mass points.

We assume that the uncertainties in jet energy scale, jet energy resolution, $\PQb$-tagging scale factors, and the total
integrated luminosity are correlated between the two samples. All other systematic uncertainties are assumed to be
uncorrelated. These include the renormalization and factorization scale and $\pt$ reweighting uncertainties, since different generators are used
for the simulation of $\ttbar$ events and the hadronic channel extracts the $\ttbar$ normalization from data.  The method for setting combined cross section upper limits is identical to the limit setting procedure of the
all hadronic channel, except a joint likelihood is used that considers both channels.

\begin{figure}[htb]
\centering
\includegraphics[width=0.7\textwidth]{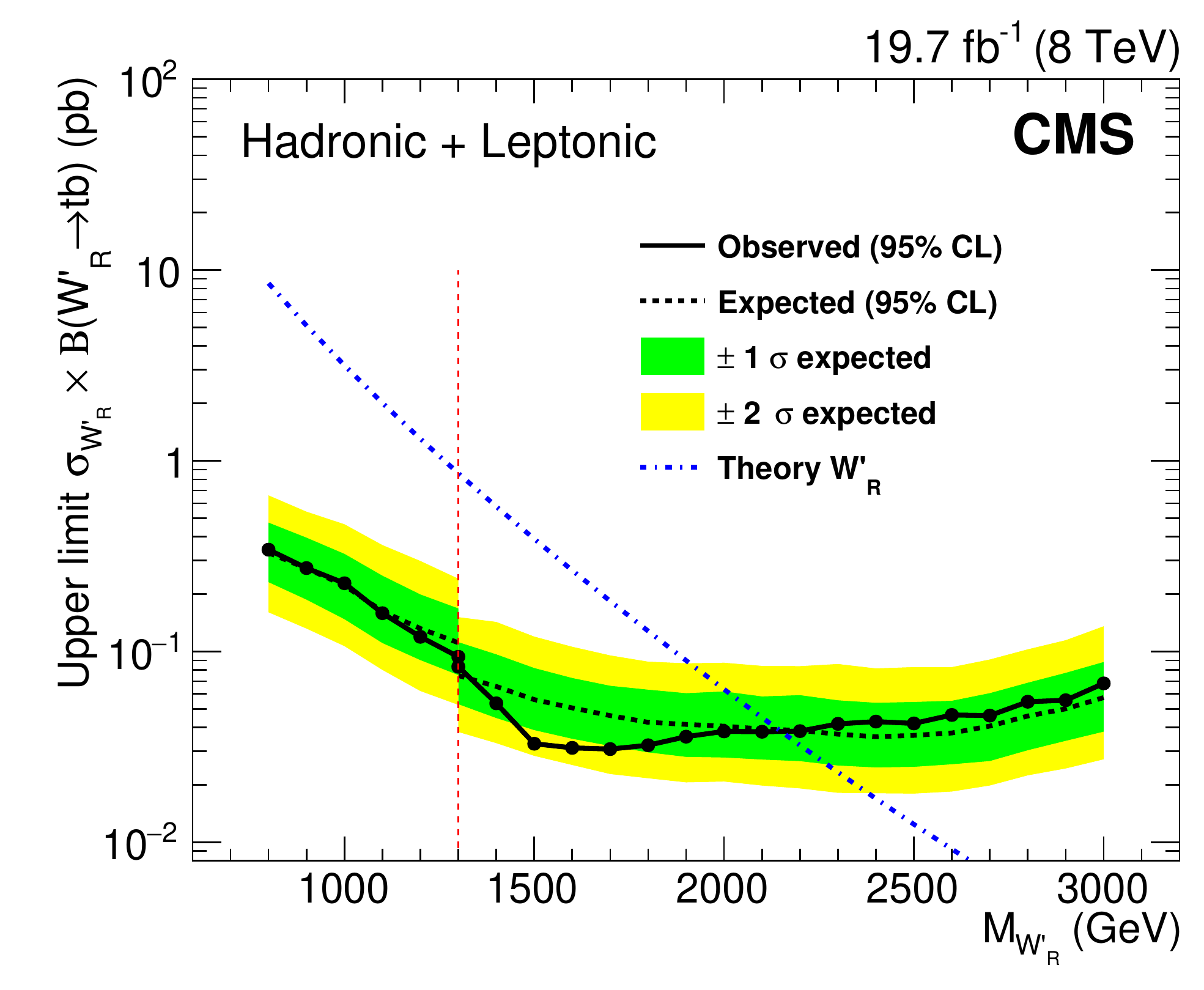}
\caption{The $\wprr$ boson 95\% CL production cross section times $\mathcal{B}_{\mathrm{\wprr \rightarrow tb}}$ limits for the combined hadronic and leptonic channels.  The observed (solid) and expected (dashed) limits, as well as the $\wprr$ boson theoretical cross section (dash-dotted) are shown.
As indicated in the legend, the shaded regions about the expected limits represent 1 and 2$\sigma$ bands.  The region to the left of the vertical dashed line shows the limits derived only from the leptonic channel.  The right of the vertical dashed line shows limits based on the combined hadronic and leptonic channels. }
\label{figs:thetalimitcombo}
\end{figure}

The $\wprr$ combined cross section upper limits are shown in Fig. \ref{figs:thetalimitcombo}.
A $\wprr$ boson with a mass below 2.15\TeV is excluded at 95\% CL.

\subsection{Generalized coupling limits}
\label{sec:GCTheta}
To set limits on generic couplings, we use the procedure outlined in Ref. \cite{Chatrchyan:2014koa}.
Because the left-handed and mixed-coupling samples cannot be separated from SM single top quark production,
we set limits on the couplings $\mathrm{a^L}$ and $\mathrm{a^R}$.
The SM single top quark production is negligible when setting cross section limits.
As it is used for signal modeling in the generalized coupling analysis, it needs to be included in that calculation.
For limit setting, we reweight the signal templates (single top quark, $\wprr$, $\wprl$, $\wprlr$) to form a combined signal template with the following cross section:
\begin{equation}\begin{split}\label{eq:xsec}
\sigma_{a^{\mathrm{L}}_{\mathrm{ud}},a^{\mathrm{R}}_{\mathrm{ud}},a^{\mathrm{L}}_{\mathrm{tb}},a^{\mathrm{R}}_{\mathrm{tb}}} = \left(1-a^{\mathrm{L}}_{\mathrm{ud}} a^{\mathrm{L}}_{\mathrm{tb}}\right){\sigma}_{\text{t}}  &+
a^{\mathrm{R}}_{\mathrm{ud}} a^{\mathrm{R}}_{\mathrm{tb}}\frac{a^{\mathrm{R}}_{\mathrm{ud}} a^{\mathrm{R}}_{\mathrm{tb}} - a^{\mathrm{L}}_{\mathrm{ud}} a^{\mathrm{L}}_{\mathrm{tb}}}{a^{\mathrm{L}}_{\mathrm{ud}} a^{\mathrm{L}}_{\mathrm{tb}} + a^{\mathrm{R}}_{\mathrm{ud}} a^{\mathrm{R}}_{\mathrm{tb}}} {\sigma}_{\wprr} \\
&+a^{\mathrm{L}}_{\mathrm{ud}} a^{\mathrm{L}}_{\mathrm{tb}}\frac{a^{\mathrm{L}}_{\mathrm{ud}} a^{\mathrm{L}}_{\mathrm{tb}} - a^{\mathrm{R}}_{\mathrm{ud}} a^{\mathrm{R}}_{\mathrm{tb}}}{a^{\mathrm{L}}_{\mathrm{ud}} a^{\mathrm{L}}_{\mathrm{tb}} + a^{\mathrm{R}}_{\mathrm{ud}} a^{\mathrm{R}}_{\mathrm{tb}}} {\sigma}_{\wprl} +
 2\frac{a^{\mathrm{R}}_{\mathrm{ud}} a^{\mathrm{R}}_{\mathrm{tb}} a^{\mathrm{L}}_{\mathrm{ud}} a^{\mathrm{L}}_{\mathrm{tb}}}{a^{\mathrm{L}}_{\mathrm{ud}} a^{\mathrm{L}}_{\mathrm{tb}} + a^{\mathrm{R}}_{\mathrm{ud}} a^{\mathrm{R}}_{\mathrm{tb}}} {\sigma}_{\wprlr},
\end{split}\end{equation}
where ${\sigma}_{\wprr}$, ${\sigma}_{\wprl}$, ${\sigma}_{\wprlr}$ are the cross section for right-handed, left-handed, or mixed samples,
respectively and $\sigma_{\text{t}}$ is the
SM $s$-channel single top quark production cross section.
We assume that the left-handed and right-handed coupling constants are equal for first and third generation quarks ($a^{\mathrm{L}}_{\mathrm{ud}} = a^{\mathrm{L}}_{\mathrm{tb}}$ and $a^{\mathrm{R}}_{\mathrm{ud}} = a^{\mathrm{R}}_{\mathrm{tb}}$).

The templates are then summed and cross section upper limits are set
using the resultant yield as the signal process for the given values of $a^{\mathrm{L}}$ and $a^{\mathrm{R}}$.
Limits are calculated using pairwise combinations of the couplings from 0 to 1 in increments of 0.1.

\begin{figure}[htbp]
\centering
\includegraphics[width=0.7\textwidth]{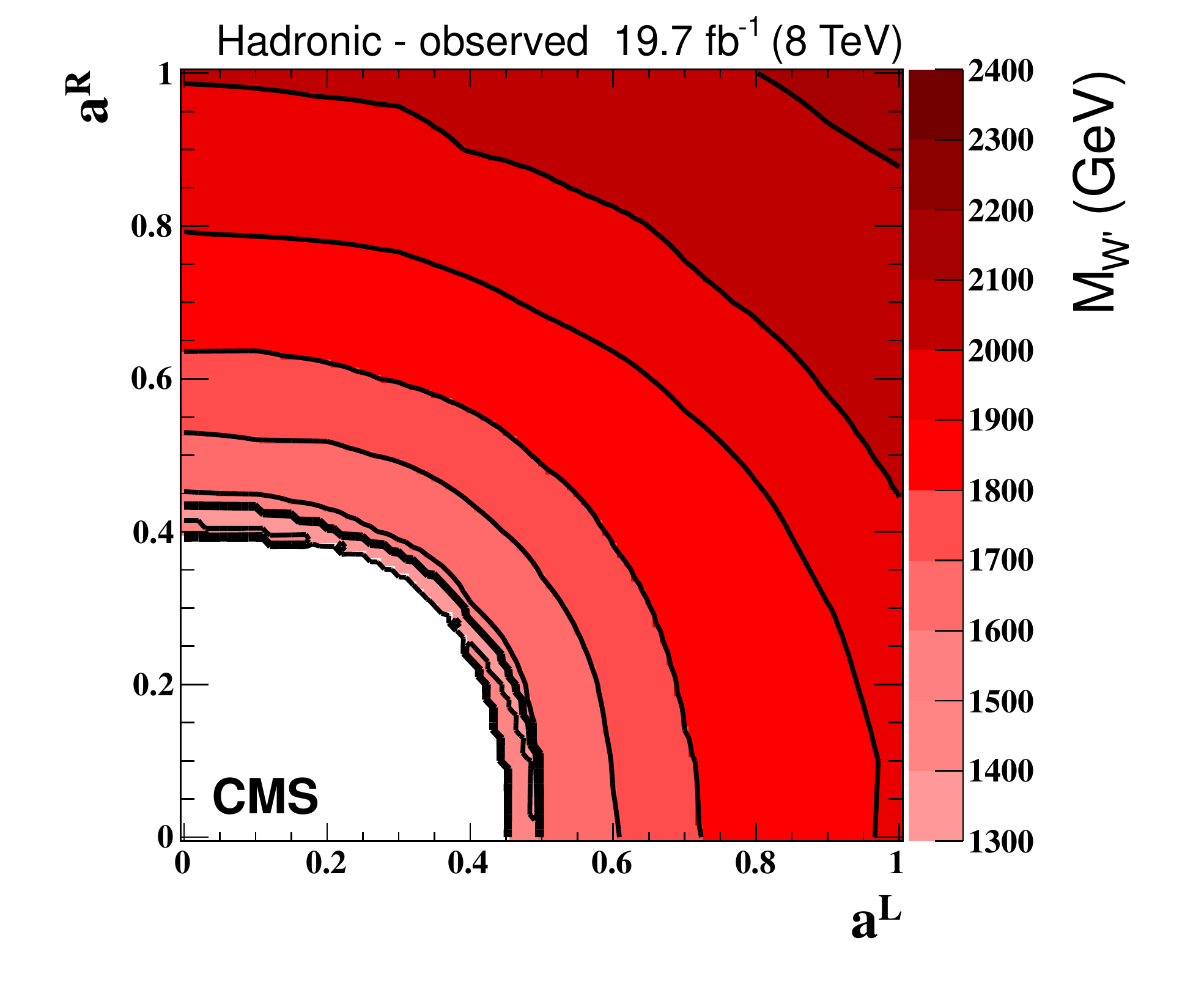}
\includegraphics[width=0.7\textwidth]{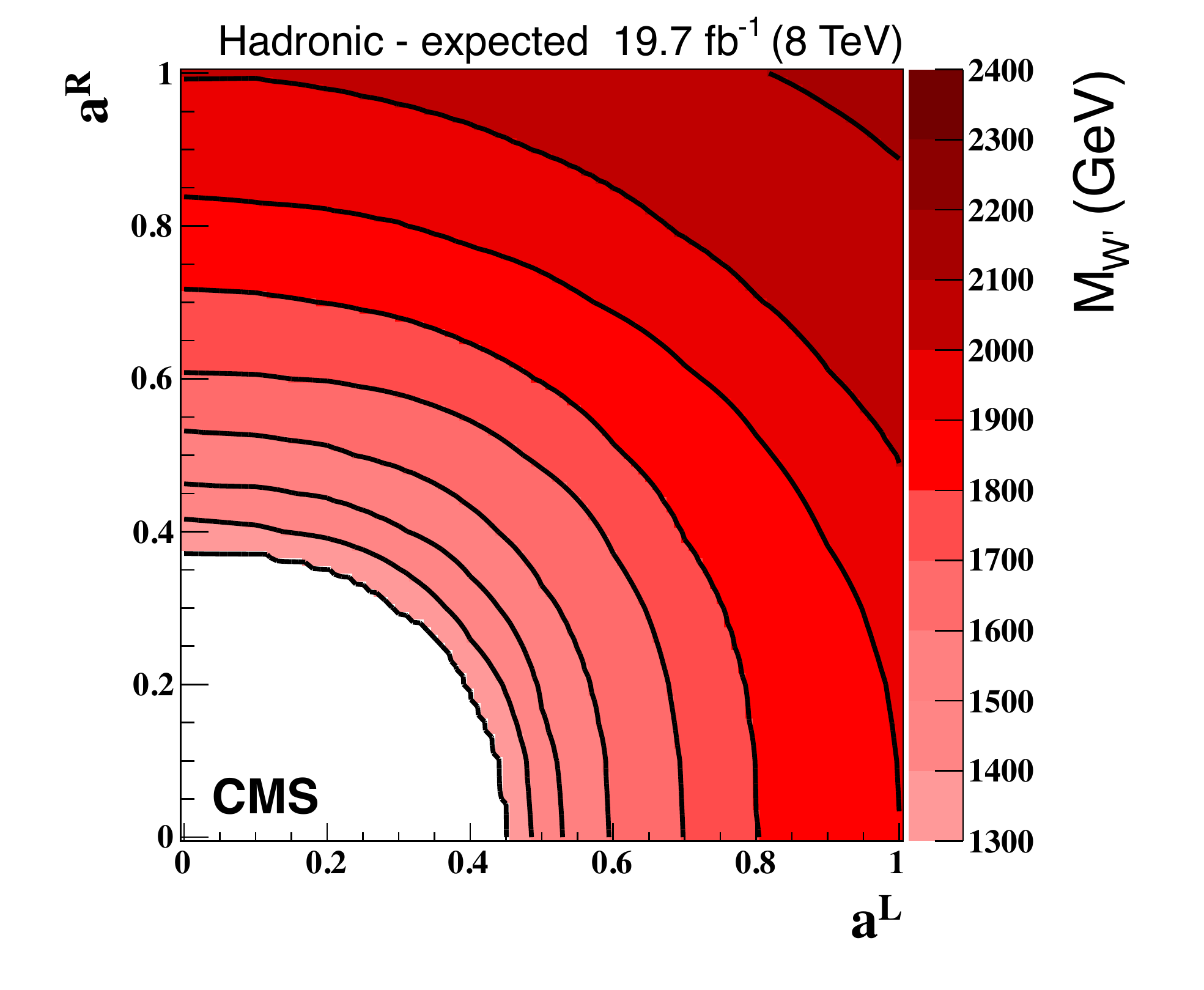}
\caption{
Contour plots of $M_{\wpr}$ in the ($a^{\mathrm{L}},~a^{\mathrm{R}}$) plane in the hadronic channel.  The top (bottom) plot shows observed (expected) limits.  The contour shading indicates the values of $M_{\wpr}$ where the theoretical cross section is equal
to the observed or expected 95\% CL limit.  }
\label{figs:GCLimhad}
\end{figure}

Using these cross section upper limits, we obtain the values where the $M_{\wpr}$ cross section limits are equal
to the theoretical cross section prediction for a given combination of $a^{\mathrm{L}}$ and $a^{\mathrm{R}}$.
These points are the maximum excluded $\wpr$ boson mass for given values of $a^{\mathrm{L}}$ and $a^{\mathrm{R}}$.
Exclusion limits when considering generalized couplings can therefore be represented with contours on a two dimensional plot in
$a^{\mathrm{L}}$ and $a^{\mathrm{R}}$ and the contour is the maximum excluded $\wpr$ boson mass.
These contours in the ($a^{\mathrm{L}},~a^{\mathrm{R}}$) plane are shown in Fig.~\ref{figs:GCLimhad} for both observed and expected limits
in the hadronic channel.
The mass upper limits for left-handed and mixed-coupling $\wpr$ bosons are 1.92 and 2.15$\TeV$, respectively.

For this procedure, no systematic uncertainty is considered from single top quark production
since the templates are dominated by statistical uncertainties.

\begin{figure}[htbp]
\centering
\includegraphics[width=0.7\textwidth]{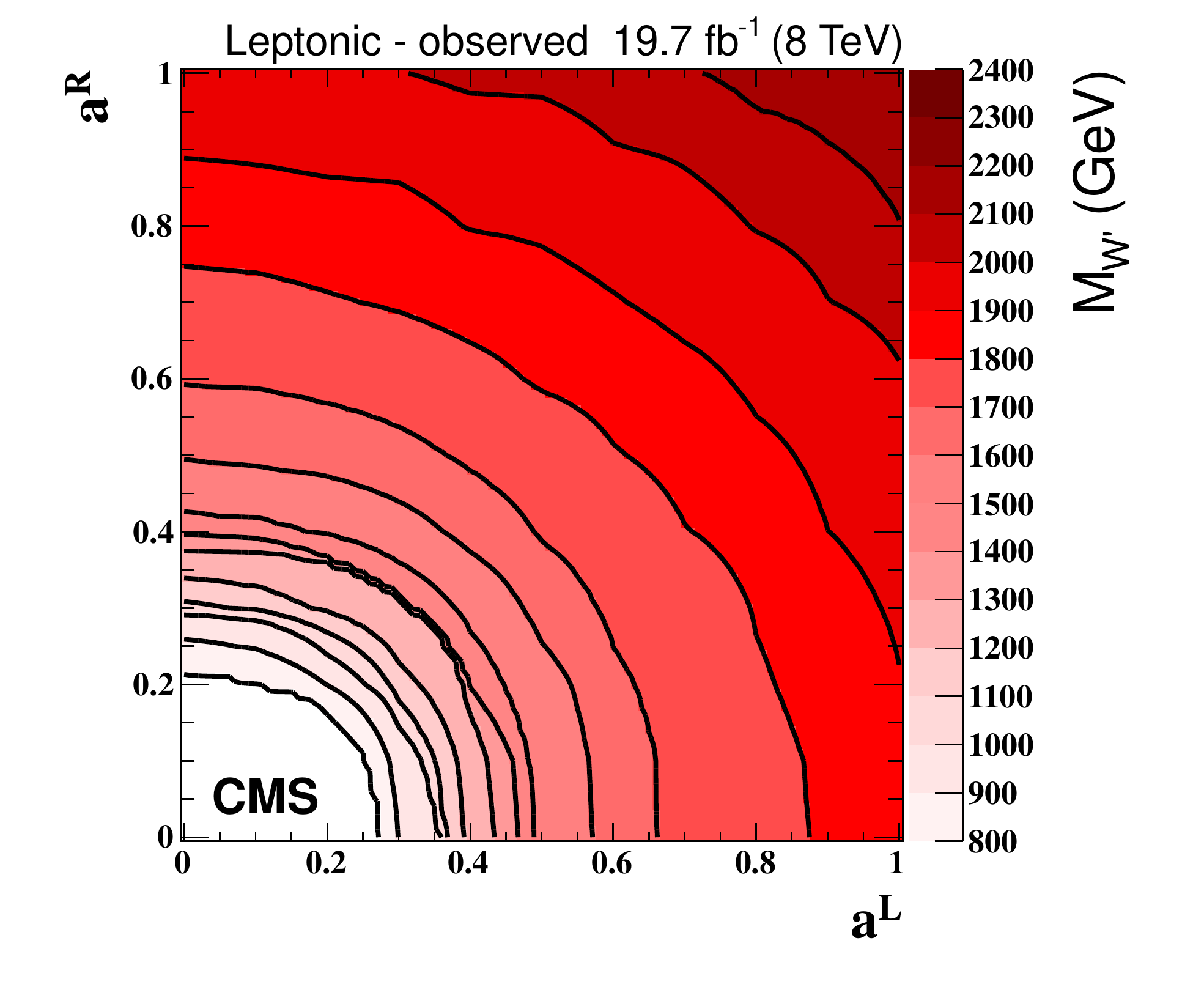}
\includegraphics[width=0.7\textwidth]{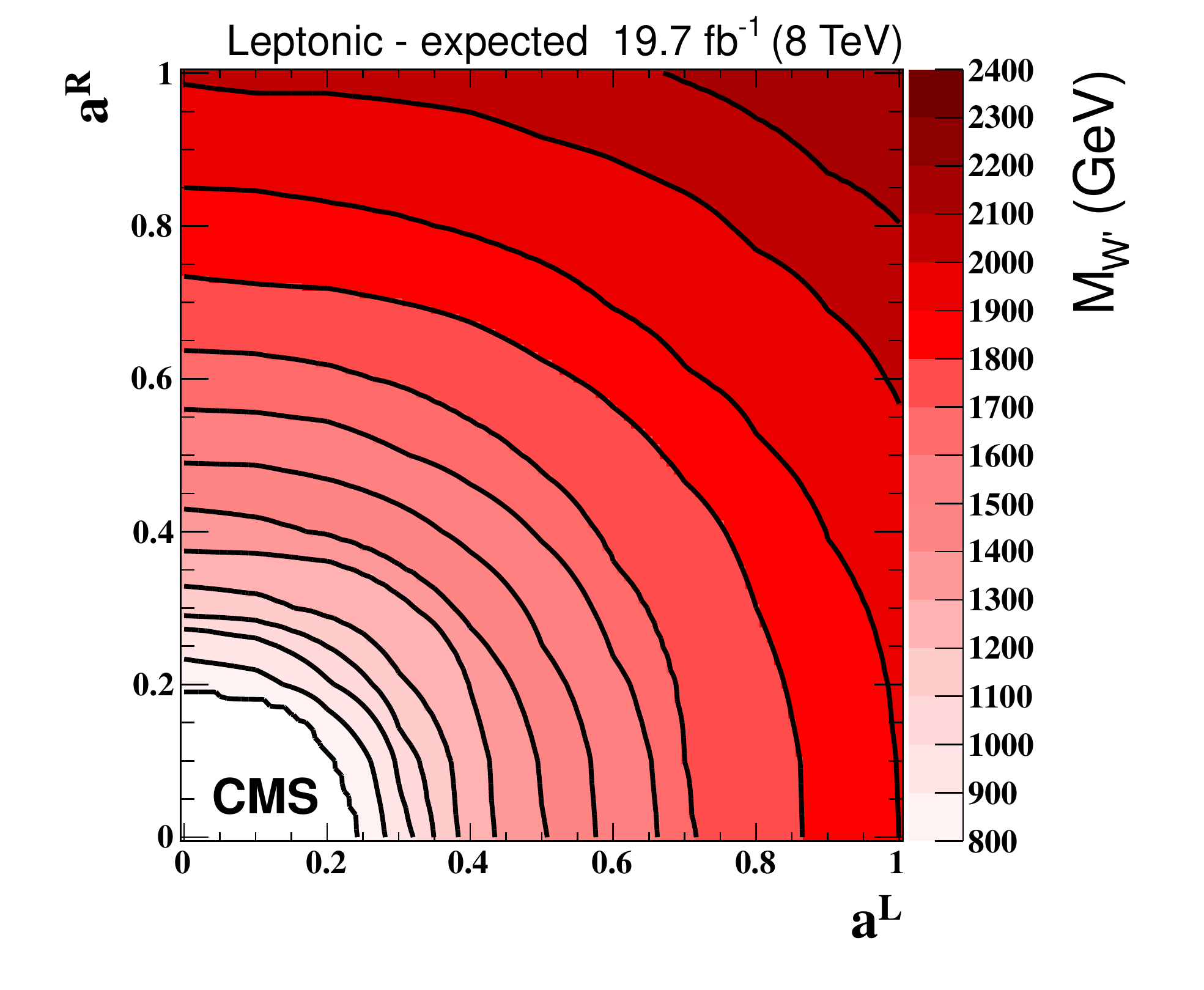}
\caption{
Contour plots of $M_{\wpr}$ in the ($a^{\mathrm{L}},~a^{\mathrm{R}}$) plane in the leptonic channel.  The top (bottom) plot shows observed (expected) limits.  The contour shading indicates the values of $M_{\wpr}$ where the theoretical cross section is equal
to the observed or expected 95\% CL limit.  These are updated from Ref.~\cite{Chatrchyan:2014koa}
to include the effect of the $\wpr$ width on the cross section of an $a^{\mathrm{L}}$ and $a^{\mathrm{R}}$ combination. }
\label{figs:GCLimlep}
\end{figure}

\begin{figure}[htbp]
\centering
\includegraphics[width=0.7\textwidth]{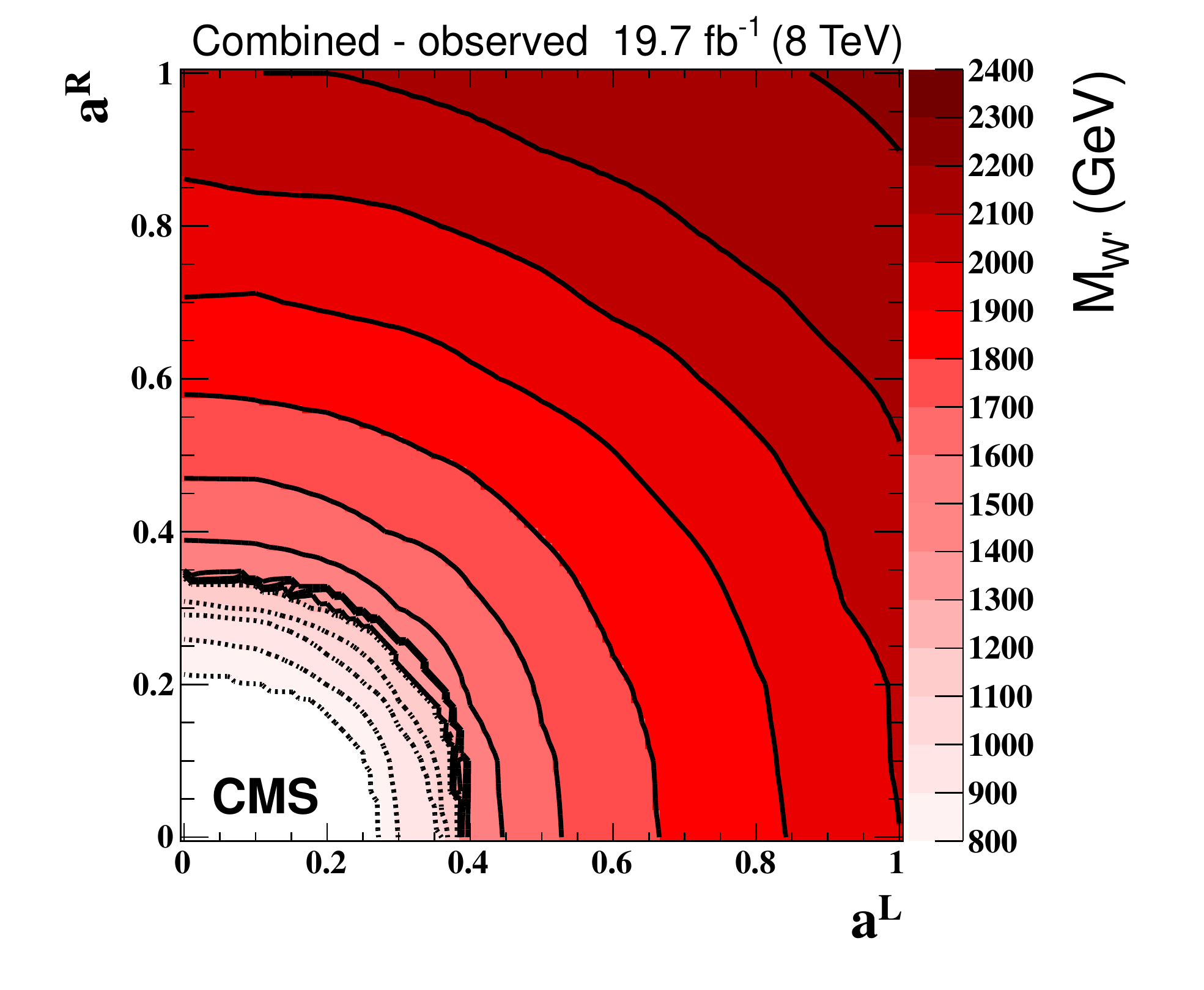}
\includegraphics[width=0.7\textwidth]{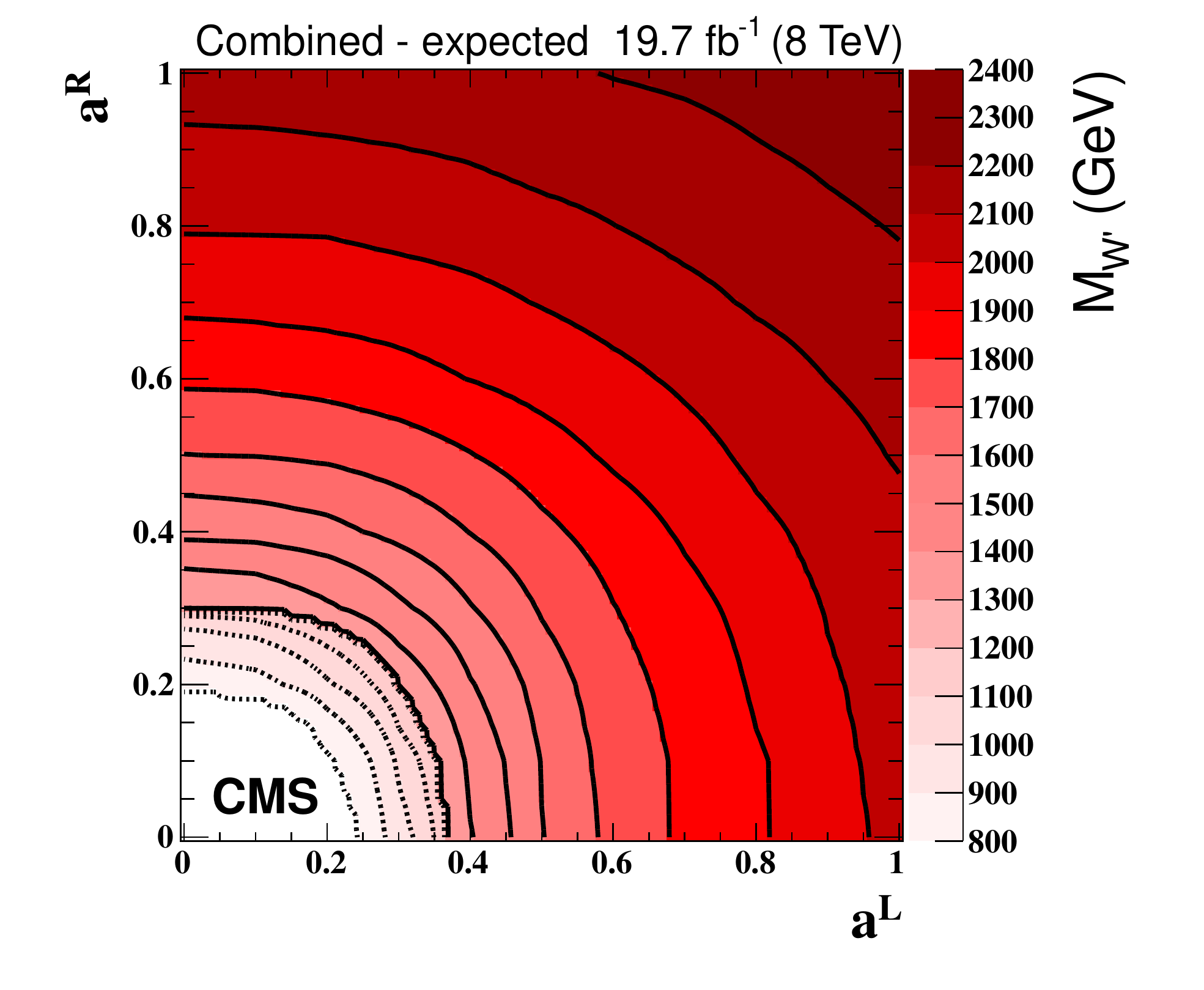}
\caption{Contour plots of $M_{\wpr}$ in the ($a^{\mathrm{L}},~a^{\mathrm{R}}$) plane using the combined hadronic and leptonic channels.  The top (bottom) plot shows observed (expected) limits.
The contour shading indicates the values of $M_{\wpr}$ where the theoretical cross section intersects the observed or expected limit band.
The solid contours represent combined limits, and the dashed contours indicate limits that are obtained purely from the leptonic channel.
}
\label{figs:GCLimcombo}
\end{figure}

Additionally, we present combined limits on the $\wpr$ coupling strengths, $a^{\mathrm{L}}$ and $a^{\mathrm{R}}$.
The limits for the leptonic channel that are used for the combination are shown in Fig.~\ref{figs:GCLimlep}.  These are updated from Ref.~\cite{Chatrchyan:2014koa}
to include the effect of the $\wpr$ width on the cross section of an $a^{\mathrm{L}}$ and $a^{\mathrm{R}}$ combination.
The limits for the combined hadronic and leptonic channels are shown in Fig.~\ref{figs:GCLimcombo}.

\section{Summary}
\label{sec:summary}

A search for a new massive gauge boson $\wpr$ decaying to a top and a bottom quark with a
hadronic signature has been performed using proton-proton collisions recorded by the CMS
detector at $\sqrt{s}=8\TeV$, corresponding to an integrated luminosity of 19.7\fbinv.
The search is focused on $\wpr$ masses above 1.30$\TeV$, where the main feature of this topology is a top quark whose decay products merge into a single jet.

The analysis uses jet substructure algorithms to distinguish the
top quark jet from standard model hadronic jet backgrounds. The principal
background, from QCD multijet production, is estimated from data using the average $\PQb$-tagging rate
measured in a QCD enhanced control region. The other important source of background, from standard model
$\ttbar$ production, is estimated from simulation, and is corrected by a scale factor derived from control samples in data.

Limits are placed on the production cross section of a right-handed $\wpr$ boson, together with constraints
on the left-handed and right-handed couplings of the $\wpr$ boson to quarks. The production of a right-handed $\wpr$ boson with
a mass below $2.02\TeV$ decaying to a hadronic final state is excluded at 95\% confidence level.  The lower limit on the mass of the right-handed
$\wpr$ boson increases to $2.15\TeV$ at 95\% confidence level when both hadronic and leptonic decays are considered, and is
the most stringent lower mass limit to date in the tb decay mode.

\section*{Acknowledgements}

\hyphenation{Bundes-ministerium Forschungs-gemeinschaft Forschungs-zentren} We congratulate our colleagues in the CERN accelerator departments for the excellent performance of the LHC and thank the technical and administrative staffs at CERN and at other CMS institutes for their contributions to the success of the CMS effort. In addition, we gratefully acknowledge the computing centres and personnel of the Worldwide LHC Computing Grid for delivering so effectively the computing infrastructure essential to our analyses. Finally, we acknowledge the enduring support for the construction and operation of the LHC and the CMS detector provided by the following funding agencies: the Austrian Federal Ministry of Science, Research and Economy and the Austrian Science Fund; the Belgian Fonds de la Recherche Scientifique, and Fonds voor Wetenschappelijk Onderzoek; the Brazilian Funding Agencies (CNPq, CAPES, FAPERJ, and FAPESP); the Bulgarian Ministry of Education and Science; CERN; the Chinese Academy of Sciences, Ministry of Science and Technology, and National Natural Science Foundation of China; the Colombian Funding Agency (COLCIENCIAS); the Croatian Ministry of Science, Education and Sport, and the Croatian Science Foundation; the Research Promotion Foundation, Cyprus; the Ministry of Education and Research, Estonian Research Council via IUT23-4 and IUT23-6 and European Regional Development Fund, Estonia; the Academy of Finland, Finnish Ministry of Education and Culture, and Helsinki Institute of Physics; the Institut National de Physique Nucl\'eaire et de Physique des Particules~/~CNRS, and Commissariat \`a l'\'Energie Atomique et aux \'Energies Alternatives~/~CEA, France; the Bundesministerium f\"ur Bildung und Forschung, Deutsche Forschungsgemeinschaft, and Helmholtz-Gemeinschaft Deutscher Forschungszentren, Germany; the General Secretariat for Research and Technology, Greece; the National Scientific Research Foundation, and National Innovation Office, Hungary; the Department of Atomic Energy and the Department of Science and Technology, India; the Institute for Studies in Theoretical Physics and Mathematics, Iran; the Science Foundation, Ireland; the Istituto Nazionale di Fisica Nucleare, Italy; the Ministry of Science, ICT and Future Planning, and National Research Foundation (NRF), Republic of Korea; the Lithuanian Academy of Sciences; the Ministry of Education, and University of Malaya (Malaysia); the Mexican Funding Agencies (CINVESTAV, CONACYT, SEP, and UASLP-FAI); the Ministry of Business, Innovation and Employment, New Zealand; the Pakistan Atomic Energy Commission; the Ministry of Science and Higher Education and the National Science Centre, Poland; the Funda\c{c}\~ao para a Ci\^encia e a Tecnologia, Portugal; JINR, Dubna; the Ministry of Education and Science of the Russian Federation, the Federal Agency of Atomic Energy of the Russian Federation, Russian Academy of Sciences, and the Russian Foundation for Basic Research; the Ministry of Education, Science and Technological Development of Serbia; the Secretar\'{\i}a de Estado de Investigaci\'on, Desarrollo e Innovaci\'on and Programa Consolider-Ingenio 2010, Spain; the Swiss Funding Agencies (ETH Board, ETH Zurich, PSI, SNF, UniZH, Canton Zurich, and SER); the Ministry of Science and Technology, Taipei; the Thailand Center of Excellence in Physics, the Institute for the Promotion of Teaching Science and Technology of Thailand, Special Task Force for Activating Research and the National Science and Technology Development Agency of Thailand; the Scientific and Technical Research Council of Turkey, and Turkish Atomic Energy Authority; the National Academy of Sciences of Ukraine, and State Fund for Fundamental Researches, Ukraine; the Science and Technology Facilities Council, UK; the US Department of Energy, and the US National Science Foundation.

Individuals have received support from the Marie-Curie programme and the European Research Council and EPLANET (European Union); the Leventis Foundation; the A. P. Sloan Foundation; the Alexander von Humboldt Foundation; the Belgian Federal Science Policy Office; the Fonds pour la Formation \`a la Recherche dans l'Industrie et dans l'Agriculture (FRIA-Belgium); the Agentschap voor Innovatie door Wetenschap en Technologie (IWT-Belgium); the Ministry of Education, Youth and Sports (MEYS) of the Czech Republic; the Council of Science and Industrial Research, India; the HOMING PLUS programme of the Foundation for Polish Science, cofinanced from European Union, Regional Development Fund; the OPUS programme of the National Science Center (Poland); the Compagnia di San Paolo (Torino); the Consorzio per la Fisica (Trieste); MIUR project 20108T4XTM (Italy); the Thalis and Aristeia programmes cofinanced by EU-ESF and the Greek NSRF; the National Priorities Research Program by Qatar National Research Fund; the Rachadapisek Sompot Fund for Postdoctoral Fellowship, Chulalongkorn University (Thailand); and the Welch Foundation, contract C-1845.

\bibliography{auto_generated}

\cleardoublepage \appendix\section{The CMS Collaboration \label{app:collab}}\begin{sloppypar}\hyphenpenalty=5000\widowpenalty=500\clubpenalty=5000\textbf{Yerevan Physics Institute,  Yerevan,  Armenia}\\*[0pt]
V.~Khachatryan, A.M.~Sirunyan, A.~Tumasyan
\vskip\cmsinstskip
\textbf{Institut f\"{u}r Hochenergiephysik der OeAW,  Wien,  Austria}\\*[0pt]
W.~Adam, E.~Asilar, T.~Bergauer, J.~Brandstetter, E.~Brondolin, M.~Dragicevic, J.~Er\"{o}, M.~Flechl, M.~Friedl, R.~Fr\"{u}hwirth\cmsAuthorMark{1}, V.M.~Ghete, C.~Hartl, N.~H\"{o}rmann, J.~Hrubec, M.~Jeitler\cmsAuthorMark{1}, V.~Kn\"{u}nz, A.~K\"{o}nig, M.~Krammer\cmsAuthorMark{1}, I.~Kr\"{a}tschmer, D.~Liko, I.~Mikulec, D.~Rabady\cmsAuthorMark{2}, B.~Rahbaran, H.~Rohringer, J.~Schieck\cmsAuthorMark{1}, R.~Sch\"{o}fbeck, J.~Strauss, W.~Treberer-Treberspurg, W.~Waltenberger, C.-E.~Wulz\cmsAuthorMark{1}
% ...
\vskip\cmsinstskip
% ...
1:~~Also at Vienna University of Technology, Vienna, Austria\\
2:~~Also at CERN, European Organization for Nuclear Research, Geneva, Switzerland\\
% ...

\end{sloppypar}
\end{document}